\PassOptionsToPackage{dvipsnames}{xcolor}
\documentclass[acmsmall,screen]{acmart}
\AtBeginDocument{%
  }

\setcopyright{acmlicensed}
\copyrightyear{2018}
\acmYear{2018}
\acmDOI{XXXXXXX.XXXXXXX}

\usepackage{paralist}
\usepackage{xspace}
\usepackage{lipsum}
\usepackage{hyperref}
\usepackage{listings}
\usepackage[dvipsnames]{xcolor}

\usepackage{multirow}
\usepackage{array}
\usepackage{booktabs}

\usepackage{hyperref}
\usepackage{ifthen}
\usepackage[most]{tcolorbox}
\usepackage{fontawesome5}
\usepackage{enumitem}
\usepackage{subcaption}
\usepackage{hyperref}
\usepackage{pifont}
\usepackage{longtable}
\usepackage{tablefootnote}

\usepackage[]{mdframed}
\usepackage{colortbl}
\usepackage{array}
\usepackage{graphicx}

\definecolor{rqboxcolor}{HTML}{f2f2ff}
\definecolor{darkgreen}{HTML}{009B55}

\newcommand{\tick}[0]{\ding{51}}

\newcommand{\totalPapers}{71\xspace}
\newcommand{\totalStudies}{67\xspace}

\newcommand{\entrypoint}[0]{$EP$}
\newcommand{\method}[0]{$MF$}
\newcommand{\fragment}[0]{$CF$}

\newcommand{\rotheader}[1]{\rotatebox{90}{\textbf{#1}}}

\newcommand{\rqOne}{What \textbf{type of software and type of vulnerabilities} are targeted by APG approaches?}
\newcommand{\rqTwo}{What \textbf{kind of artifact} (i.e., probe) is generated by APG approaches?}
\newcommand{\rqThree}{What \textbf{techniques and tools} are adopted by APG approaches to generate probes?}
\newcommand{\rqFour}{What \textbf{kind of input} is required by APG approaches to be started?}
\newcommand{\rqFive}{How has the overall \textbf{effectiveness} of APG approaches been \textbf{evaluated}?}
\newcommand{\rqSix}{How are APG approaches \textbf{implemented} and \textbf{released}?}

\mdfdefinestyle{boxstyle1}{%
    linecolor=black,
    linewidth=2pt,%
    leftmargin=0cm,rightmargin=0cm,%
    roundcorner=5pt,
    topline=false,bottomline=false,rightline=false,%
    backgroundcolor=black!10
}
\mdfdefinestyle{boxstyle2}{%
    linecolor=purple,linewidth=2pt,%
    leftmargin=0cm,rightmargin=0cm,%
    roundcorner=5pt,
    topline=true,bottomline=true,rightline=false,leftline=false,%
    backgroundcolor=black!05
}
\mdfdefinestyle{boxstyle3}{%
    linecolor=purple,linewidth=2pt,%
    leftmargin=0cm,rightmargin=0cm,%
    roundcorner=5pt,
    topline=false,bottomline=false,rightline=true,leftline=true,%
    backgroundcolor=purple!10
}
\mdfdefinestyle{boxstyle4}{%
    linecolor=black,linewidth=2pt,%
    leftmargin=0cm,rightmargin=0cm,%
    roundcorner=5pt,
    topline=true,bottomline=true,rightline=false,leftline=false,%
    backgroundcolor=black!10,
    frametitlebackgroundcolor=black
}
\mdfdefinestyle{boxstyle5}{%
    linecolor=darkgreen,
    linewidth=2pt,%
    leftmargin=0cm,rightmargin=0cm,%
    roundcorner=5pt,
    topline=false,bottomline=false,rightline=true,leftline=true,%
    backgroundcolor=darkgreen!10
}

\newcommand{\newrqbox}[2]
{
    \medskip
    \begin{mdframed}[style=boxstyle1]
    {\textbf{\faSearch\ RQ$_{#1}$.} \textit{#2}}
    \end{mdframed}
    \smallskip
}
\newcommand{\discussion}[2]
{
    \medskip
    \begin{mdframed}[style=boxstyle3]
    {\faLightbulb\ \underline{\textbf{#1}} #2}
    \end{mdframed}
    \smallskip
}
\newcommand{\rqanswer}[2]
{
    \smallskip
    \begin{mdframed}[style=boxstyle5]
    {\textbf{\faHandPointRight\ Answer to RQ$_{#1}$.} #2}
    \end{mdframed}
}

\makeatletter
\newcommand{\labelstudy}[1]{%
	\phantomsection
	\renewcommand{\@currentlabel}{#1}
	\label{#1}
}
\makeatother

\begin{document}

\title{Automated Exploit and Security Test Generation: A Systematic Literature Review}

\author{Quang-Cuong Bui}
\orcid{0000-0001-6072-9213}
\affiliation{%
  \institution{Hamburg University of Technology}
  \city{Hamburg}
  \country{Germany}
}
\email{cuong.bui@tuhh.de}

\author{Emanuele Iannone}
\authornote{Corresponding Author}
\orcid{0000-0001-7489-9969}
\affiliation{%
  \institution{Hamburg University of Technology}
  \city{Hamburg}
  \country{Germany}
}
\email{emanuele.iannone@tuhh.de}

\author{Maria Camporese}
\orcid{0009-0009-1178-0210}
\affiliation{%
  \institution{University of Trento}
  \city{Trento}
  \country{Italy}
}
\email{maria.camporese@unitn.it}

\author{Torge Hinrichs}
\orcid{0000-0001-7489-3540}
\affiliation{%
  \institution{Hamburg University of Technology}
  \city{Hamburg}
  \country{Germany}
}
\email{torge.hinrichs@tuhh.de}

\author{Catherine Tony}
\orcid{0000-0002-9916-4456}
\affiliation{%
  \institution{Hamburg University of Technology}
  \city{Hamburg}
  \country{Germany}
}
\email{catherine.tony@tuhh.de}

\author{László Tóth}
\orcid{0000-0002-5797-8670}
\affiliation{%
  \institution{University of Szeged}
  \city{Szeged}
  \country{Hungary}
}
\email{premissa@inf.u-szeged.hu}

\author{Fabio Palomba}
\orcid{0000-0001-9337-5116}
\affiliation{%
  \institution{University of Salerno}
  \city{Salerno}
  \country{Italy}
}
\email{fpalomba@unisa.it}

\author{Péter Hegedűs}
\orcid{0000-0003-4592-6504}
\affiliation{%
  \institution{University of Szeged, FrontEndART Ltd.}
  \city{Szeged}
  \country{Hungary}
}
\email{hpeter@inf.u-szeged.hu}

\author{Fabio Massacci}
\orcid{0000-0002-1091-8486}
\affiliation{%
  \institution{University of Trento}
  \city{Trento}
  \country{Italy}
}
\affiliation{%
  \institution{Vrije Universitiet Amsterdam}
  \city{Amsterdam}
  \country{The Netherlands}
}
\email{fabio.massacci@ieee.org}

\author{Riccardo Scandariato}
\orcid{0000-0003-3591-7671}
\affiliation{%
  \institution{Hamburg University of Technology}
  \city{Hamburg}
  \country{Germany}
}
\email{riccardo.scandariato@tuhh.de}

\renewcommand{\shortauthors}{Bui et al.}

\begin{abstract}
Static analysis techniques are often insufficient to determine whether a reported security vulnerability is exploitable and actually poses a real security risk to a software application.
Dynamic analysis addresses this limitation by executing the target application with carefully crafted inputs that trigger or reproduce a given vulnerability.
Over the past two decades, \textit{automatic exploit generation} (AEG) and \textit{automatic security test generation} (ASTG) have emerged as active research areas in cybersecurity and software engineering.

This paper demystifies the field by establishing consistent terminology and presenting a systematic literature review (SLR) of research published in the last two decades.
Starting from an initial set of 3,309 publications obtained from academic literature search engines, we selected and analyzed 71 papers comprising 67 distinct studies.
We systematically compare existing AEG and ASTG approaches with respect to (i) the targeted software and vulnerability types, (ii) the generated artifacts, (iii) the underlying generation techniques and tools employed, (iv) the required inputs, (v) the evaluation methodologies, and (vi) implementation and tool availability.
\end{abstract}




\maketitle

\section{Introduction}
\label{sec:intro}

Since 2024, the number of yearly vulnerabilities disclosed through CVE (Common Vulnerabilities and Exposures) has surpassed 40,000.\footnote{Number of CVEs by year: \url{https://www.cvedetails.com/browse-by-date.php}}
In this period, the rate of high-severity vulnerabilities, i.e., those with a CVSS (Common Vulnerability Scoring System) \textit{base score} of at least $8.0$, ranges from 28\% (in 2024 and 2025) to 34\% (in 2026, still in progress).
One notable example is Log4Shell (CVE-2021-44228)\footnote{\url{https://nvd.nist.gov/vuln/detail/CVE-2021-44228}}, a Remote Code Execution (RCE) vulnerability in Apache Log4j2 caused by a lack of sanitization of special expressions that users could inject, triggering remote JNDI lookups to load remote classes and execute their code.
Due to the widespread use of Log4j, this critical vulnerability affected a huge number of web applications; thus, it was assigned the maximum CVSSv3 score ($10.0$).

Finding vulnerabilities in software applications often relies on \textit{static approaches}, such as program analysis~\cite{livshits2005finding, bessey2010few, jovanovic2006pixy, goseva2015capability} or data-driven approaches using static code features, either manually engineered~\cite{scandariato2014predicting, du2019leopard, 5477059, walden2014predicting} or learned~\cite{li2018vuldeepecker,chakraborty2021deep,fu2022linevul, 11029911, ferrag2025securefalcon, lu2024grace, zhang2026vultrlm, yang2026context}.
However, these have been known to report several false positives: \textit{not all warnings indicate real vulnerabilities}.
Besides, even if a vulnerability really exists in the code (i.e., a true positive), \textit{not all vulnerabilities are exploitable} given the context in which they exist.

To address this problem, \textbf{dynamic approaches} send real input to the software application to trigger concrete signals indicating that a vulnerability is present, such as crashes, memory safety violations, privilege changes, information leaks, or unintended side effects~\cite{cybok-ss}.
Popular dynamic analysis tools include \textit{vulnerability scanners}, which are typically used for web applications~\cite{li:www2023:scanner}, and \textit{fuzzers}, which send the target program pseudo-random input, often guided by coverage information~\cite{Liang:tr:2018:fuzzing:slr,Chen:cose2018:fuzzing:review,Mallissery:csur2023:fuzzing,Zhao:sc2024:fuzzing:review}.
Dynamic approaches primarily aim to \textit{find vulnerabilities}, either ``known'', i.e., confirming their presence, or ``unknown'', i.e., discovering them.
Taken alone, tools like scanners and fuzzers do not immediately \textit{assess the exploitability} of the vulnerability they found (even if the triggered signals, such as crashes, suggest that an exploit may be possible).
Therefore, over the last two decades, researchers have introduced a subtype of dynamic approach that focuses on \textbf{automatically generating exploits and security tests}, often targeting specific known vulnerabilities.
The output of these generators is \textit{reusable artifacts} that security experts and software engineers can use to (i) demonstrate a vulnerability is exploitable, (ii) assess the severity of a vulnerability, and (iii) verify the correctness of the security patch removing a vulnerability.

\subsection{Terminology Adopted in This Review}

In the academic literature, we find that many approaches have been framed as \textit{Automated Exploit Generation} (AEG) and/or \textit{Automated Security Test Generation} (ASTG).
The first problem we found is terminology, as existing studies have provided different definitions of what constitutes an \textit{exploit} and a \textit{security test}; often, they do not even make these explicit.
Both refer to \textit{executable artifacts} against a target to exercise a security vulnerability, though they appear to have different goals.
In this paper, we adopt the following senses:
\begin{itemize}
    \item \textbf{Exploit}: an executable artifact that takes advantage of a security vulnerability in a software system to \textit{unambiguously demonstrate a bug is security-critical}~\cite{avgerinos2014automatic}.
    An exploit is \textit{weaponized} when it has a \textbf{harmful payload} (i.e., data or code) that delivers a concrete attack against the target; however, exploits may also have \textbf{harmless payloads} used for demonstration and assessment purposes; in such cases, the exploits are called \textit{Proof of Concept} (PoC).
    \item \textbf{Security Test Case}: an executable artifact that invokes the test subject with specific \textit{input} under certain \textit{preconditions} and verifies the presence of a security bug through implicit or explicit \textit{assertions}~\cite{iso-iec-ieee:29119-1:2022}.
    Often, tests use harmless payloads (i.e., test input data), since they are intended only to reproduce a vulnerability or confirm a patch works as intended.
    A security test can be a \textit{unit test} if it targets functions or methods (individually or in groups), or an \textit{end-to-end test} if it targets the whole software application via its external interface (e.g., through HTTP).
    A security test is also known as Proof of Vulnerability (PoV)~\cite{bui2022vul4j,garg:icst24:mutants} or a Vulnerability-witnessing Test~\cite{kang:issta2022:transfer,antal:ease25,iannone:msr2026:vuteco:sec:tests}.
\end{itemize}

Exploits with harmful payloads are primarily made by attackers, while security experts and testers use harmless payloads to produce PoCs.
Security tests, by contrast, are primarily made by internal developers using harmless payloads (making them more similar to PoC exploits).
Exploits are commonly made when the target system is deployed, while security tests may be created when the test target is partially implemented or even absent (typical in Test-Driven Development contexts).
Due to this, security tests are versioned inside the application repository and are maintained over time.
Indeed, tests are intended to be run automatically and continuously (through CI/CD) to detect security regressions.
On the other hand, exploits are mostly manually triggered.

\begin{figure}[t]
    \centering
    \begin{subfigure}{0.45\textwidth}
        \centering
        \includegraphics[width=\linewidth]{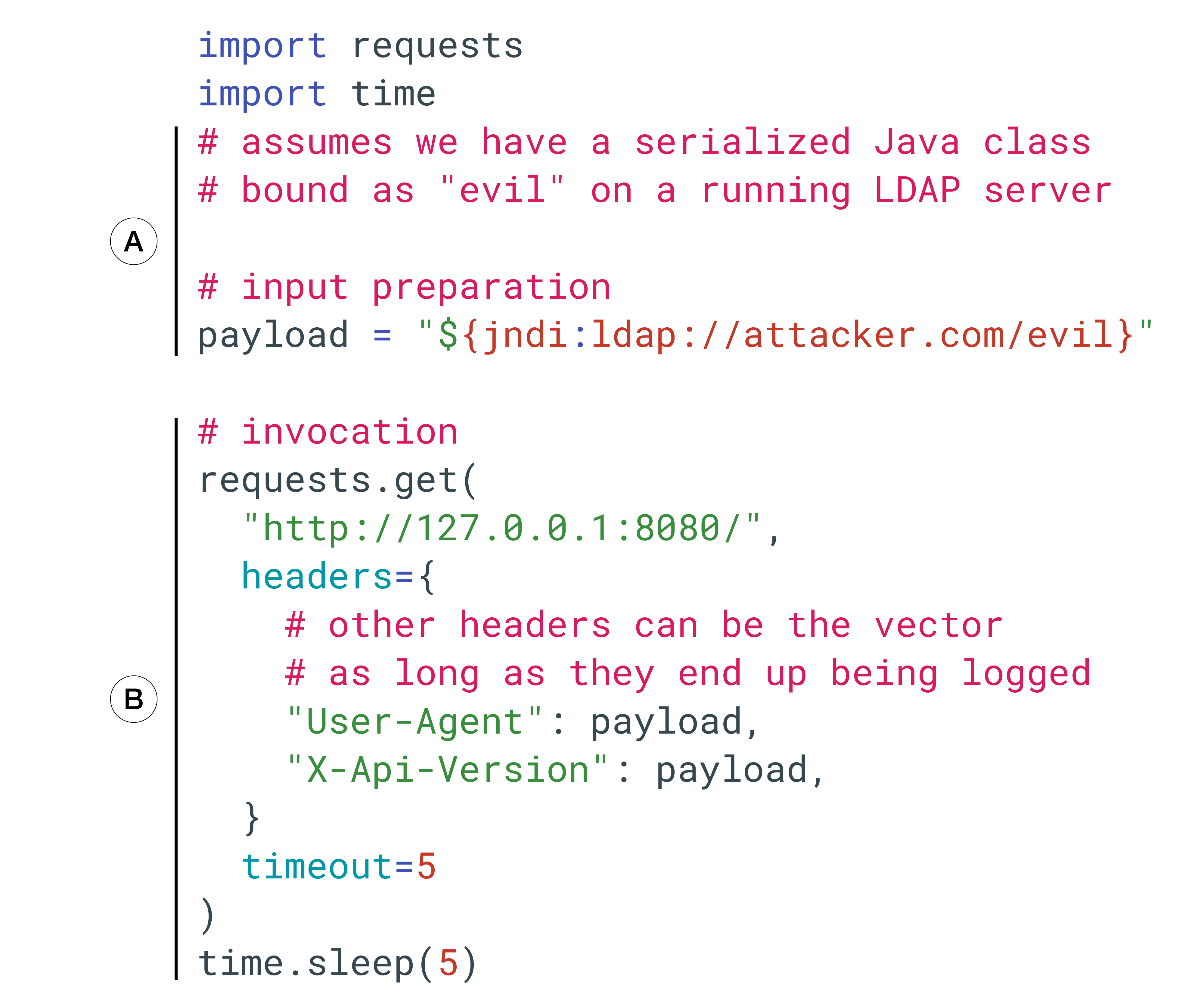}
        \caption{Example of a PoC exploit.}
        \label{fig:snippets:exploit}
    \end{subfigure}
    \hfill
    \begin{subfigure}{0.45\textwidth}
        \centering
        \includegraphics[width=\linewidth]{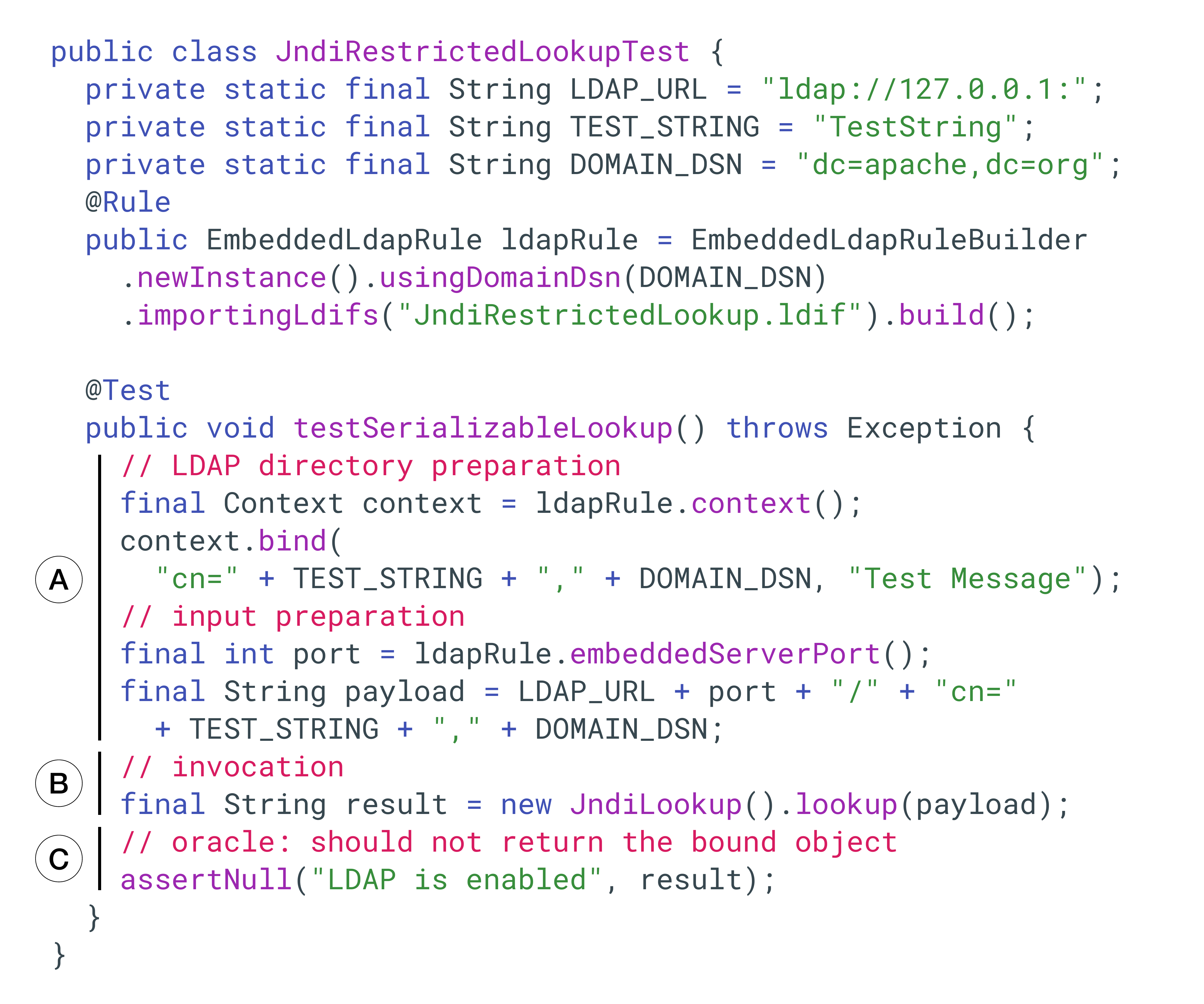}
        \caption{Example of a security test case.}
        \label{fig:snippets:test}
    \end{subfigure}
    \caption{Example of a PoC exploit and security test case for the Log4Shell vulnerability. Part \textcircled{A} prepares the payload, part \textcircled{B} invokes the target, and part \textcircled{C} confirms the presence of the vulnerability.}
    \label{fig:snippets:main}
\end{figure}

Figure~\ref{fig:snippets:main} shows an example of an exploit and a security test for Log4Shell (CVE-2021-44228).
The exploit in Figure~\ref{fig:snippets:exploit} is a simple Python script that invokes the target web application (here, at \texttt{http://127.0.0.1}) by injecting a JNDI expression into some request headers.
The exploit assumes that at least one of these headers will be logged by Log4J, and that the \texttt{\$()} expression will be evaluated to trigger a JNDI lookup.
The lookup would load the malicious Java class the attacker previously prepared on a web server under their control (here, at \texttt{ldap://attacker.com/evil}), triggering its initialization and, in turn, anything the attacker intended to run.
An exploit like this operates ``from the outside'' of the targeted application (\textit{black box}), emulating the attacker's perspective.

The test case in Figure~\ref{fig:snippets:test} is a JUnit test method that invokes the vulnerable method \texttt{lookup()} to verify that the enabled JNDI lookup restriction (the precondition) works as intended.
Specifically, the test first binds a serialized object (here, a \texttt{String} object representing any hypothetical Java class) to a local LDAP server and then checks whether the lookup result is null.
Unlike the exploit in Figure~\ref{fig:snippets:exploit}, this test does not inject any \texttt{\$()} expression, as its goal is to check whether the JNDI lookup is disallowed when configured.
A test like this operates ``from the inside'' of the targeted application (\textit{white box}), requiring direct access to object construction and methods and allowing \textit{scaffolding} and, potentially, \textit{mocking} to create the right test conditions.
Indeed, this test case is a simplified version of a test case belonging to the original \texttt{apache/logging-log4j2} project.\footnote{\url{https://github.com/apache/logging-log4j2/blob/2.x/log4j-core-test/src/test/java/org/apache/logging/log4j/core/lookup/JndiRestrictedLookupTest.java}}

From a structural perspective, the two types of artifacts have both a phase where the payload is prepared (part \textcircled{A}) and the targeted asset, either the whole application or a specific method, is invoked (part \textcircled{B}).
In the exploit, the payload consists of the \texttt{\$()} expression that triggers the attack (i.e., loading the malicious Java class from a server the attacker prepared).
In the test case, the payload corresponds to the input sent to the \texttt{lookup()} method.
Part \textcircled{C} is common in test cases and is meant to confirm that the targeted asset is free of a specific security bug, e.g., by checking the return value or other side effects.
This happens through automated \textit{assertions}, which are fundamental to the test and act as an \textit{oracle}.
Deciding the right oracle is a well-known problem in software testing~\cite{barr:tse2014:oracle:problem}; however, it remains underexplored in security testing, which also affects the quality of tests generated by ASTG approaches.
Exploits, on the other hand, do not often implement a similar mechanism.
Indeed, exploits are often human-triggered and not meant to be launched repeatedly (unlike tests), and the attacker manually verifies their success, e.g., by checking the occurrence of a crash, a memory violation, or if a connection to an attacker website was established (as for the exploit in Figure~\ref{fig:snippets:exploit}).
However, if exploits are generated automatically, having an automated oracle becomes relevant, as it would simplify assessing their correctness.

For this review, we introduce the term \textbf{Dynamic Security Probe}, or simply \textit{probe}, to refer to any artifact that is used at runtime to exercise or demonstrate a security vulnerability in a software system.
This umbrella term covers the sense we provided for exploits and security tests, as well as the nuances in the academic literature.
Likewise, we incorporate AEG and ASTG under the term \textit{Automated Probe Generation} (APG) and use it to refer to both concepts collectively for simplicity.

\subsection{Paper Contribution}

This paper systematically reviews the academic literature on \textbf{AEG and ASTG} approaches (here collectively referred to as APG) targeting \textbf{software vulnerabilities}.
Our analysis focused on identifying the types of software systems and vulnerabilities that existing APG approaches support.
Then, we checked the exact type of probe that the existing approaches generate, i.e., whether they generate exploits or security tests (according to the sense we gave), and how the probe's effectiveness is determined (i.e., the oracle).
We further characterized the approaches by examining the underlying techniques used to generate the probe and the input types required to initiate generation.
We also analyze how the APG approaches have been evaluated in their original studies (e.g., the datasets used) and the degree of automation of the released tools.

Our scope is APG approaches that produce probes, or at least parts of them.
Given this, we did not consider studies on fuzzing techniques, as their natural output is a \textit{report} on how the target behaved in response to the generated inputs.
Nevertheless, APG approaches that build on fuzzers and still produce probes were included.
Besides, fuzzing has already received attention in reviews~\cite{Liang:tr:2018:fuzzing:slr,Chen:cose2018:fuzzing:review,Zhao:sc2024:fuzzing:review} and surveys~\cite{li2018fuzzing,zhu2022fuzzing,manes2019art}.
Furthermore, we focused on software-level vulnerabilities because of the large existing body of research and the authors' expertise in the field.
Hardware-level vulnerabilities and network misconfigurations are outside the scope of this work.
In summary, the main contributions of this paper are:
\begin{itemize}
	\item A systematic literature review (SLR) on academic research works on automated probe generation (AEG and ASTG) for software vulnerabilities.
    \item An analysis of the key characteristics of APG approaches.
    \item A list of APG tools that are available and usable in practice.
\end{itemize}

\section{Related Work}
\label{subsec:related}

Over the years, several secondary studies concerning exploit generation and security testing in general have been published.
Table~\ref{tbl:related-surveys} highlights the main differences between our systematic literature review and existing secondary studies, for a total of 10.
Three are non-systematic surveys: Shahriar et al.~\cite{shahriar2009automatic}, Felderer et al.~\cite{felderer2016security}, and Liu et al.~\cite{liu2019survey}.

Shahriar and Zulkernine~\cite{shahriar2009automatic} compare 20 program security testing techniques against four common vulnerability classes (buffer overflow, SQL injection, format string bug, and cross-site scripting), classifying them along seven criteria: vulnerability coverage, source of test cases, test generation method, testing level, test-case granularity, tool automation, and target applications.
They further contrast the degree of automation for each task across three testing stages (test-case generation, oracle generation, and test-case execution), concluding that most tools target web vulnerabilities (SQLi/XSS), that automation support for buffer overflows (BOF) and format string bugs is weak, and that few tools offer full end-to-end automation.

Felderer et al.~\cite{felderer2016security} provide a broad, non-systematic survey of security testing techniques, organizing them by their test basis within the secure software development life cycle into four families: model-based security testing (analysis and design of models), code-based testing and static analysis (at the source and bytecode level), penetration testing and dynamic analysis (targeting running systems), and security regression testing (employed during maintenance).
Rather than cataloging tools along measurable dimensions, they explain the conceptual foundations of each family and illustrate their combined application through a running case study of a three-tiered web business application.

Liu et al.~\cite{liu2019survey} focus specifically on cross-site scripting (XSS), first classifying XSS into DOM-based, stored, and reflected types and then explaining concrete exploitation scenarios (phishing, cookie theft, DoS/DDoS, screenshot exfiltration, and XSS worms) on a purpose-built demo website.
They then review 30 XSS detection methods, grouping them into static, dynamic, and hybrid analysis categories and comparing their strengths, weaknesses, and the XSS sub-types each can detect, before closing with prevention guidance.

These studies selected a representative set of state-of-the-art or known approaches and techniques to compare their characteristics and effectiveness on various dimensions.
These studies do not define their search and analysis methodology, suggesting they followed the authors' knowledge of the field.

A second group of secondary studies we identified adopts a systematic approach to searching for papers to review; however, neither defines explicit research questions.
Felderer et al.~\cite{felderer2016model} propose a taxonomy for model-based security testing (MBST) and use it to systematically classify 119 publications extracted from five digital libraries.
The taxonomy comprises filter criteria (model of system security, security model of the environment, and explicit test selection criteria) and evidence criteria (maturity of the evaluated system, evidence measures, and evidence level), which the authors use to characterize the state of the art and discuss open research directions in security properties, coverage criteria, and the return on investment of MBST.

Sommer et al.~\cite{sommer2023survey} survey model-based testing (MBT) and model-based security testing (MBST) approaches specifically in the automotive domain, systematically identifying 63 MBT and 29 MBST publications and comparing them along applied models, formalisms, test-selection criteria, and test-generation techniques, as well as the vehicle technologies and development stages they target.
Their aim is to assess whether the advantages of model-based testing (early, automated test generation) can be applied to automotive security testing, and, on that basis, to identify current challenges and open research areas in automotive MBST.

\begin{table}
\centering
\footnotesize
\caption{Related secondary studies on exploits and/or security testing. A dash (--) indicates a dimension that was not considered.}
\label{tbl:related-surveys}
\resizebox{1\linewidth}{!}{%
\begin{tabular}{llcc|lllccccccc}
\hline
\multirow{2}{*}{\textbf{Study}} & \multicolumn{3}{c}{\textbf{Metadata}} & \multicolumn{5}{c}{\textbf{Analyzed Dimensions}} \\
& \textit{Type} & \textit{Survey Period} & \textit{RQs} & \textit{Subject} & \textit{Target SW/Vuls.} &  \textit{Input/Output} & \textit{Automation} & \textit{Tool} \\
\hline
Shahriar and Zulkernine (2009)~\cite{shahriar2009automatic} & Survey & 2002--2009 & & ASTG & BOF, SQLi/XSS & Both & \ding{51} & \ding{51} \\
Felderer et al. (2015)~\cite{felderer2016security} & Survey & Not specified & & Security Testing & Any & \textemdash & \textemdash & \textemdash \\
Felderer et al. (2016)~\cite{felderer2016model} & Mapping & 1996--2013 & & MBST & Any & \textemdash & \textemdash & \textemdash\\
Liu et al. (2019)~\cite{liu2019survey} & Survey & Not specified & & SAST/DAST  & XSS & \textemdash & \textemdash  & \textemdash\\
Aydos et al. (2022)~\cite{aydos2022security} & Mapping & Up to 2020 & \ding{51} & SAST/DAST & OWASP 10 & \textemdash & \ding{51} &  \textemdash  \\
Alazmi et al. (2022)~\cite{alazmi2022systematic} & SLR & Not specified & \ding{51} & Web Vuln. Scanners & OWASP 10 & \textemdash  & \textemdash & \ding{51}  \\
Sommer et al. (2023)~\cite{sommer2023survey} & SLR & Up to 2021 & & MBST & Automotive & Only Output & \ding{51}& \ding{51} \\

Ahsan et al. (2024)~\cite{ahsan2024systematic} & SLR & 2001--2022 & \ding{51} & SBSST & Any & \textemdash & \textemdash & \textemdash\\
Sutter et al. (2024)~\cite{sutter2024dynamic} & SLR & 2017--2023 & \ding{51} & DAST & Android & \textemdash & \textemdash & \ding{51} \\
Elder et al. (2024)~\cite{elder2024survey} & SLR & Up to 2022 & \ding{51} & Exploit Assessment & Any & Both & \textemdash & \textemdash \\
\hline
\textbf{This work} & SLR & 2005--2024 & \ding{51} & AEG \& ASTG & Any & Both & \ding{51} & \ding{51} \\
\hline
\end{tabular}
}
\end{table}

The remaining papers formulated specific research questions.
Aydos et al.~\cite{aydos2022security} present a mapping study, while Alazmi et al.~\cite{alazmi2022systematic}, Ahsan et al.~\cite{ahsan2024systematic}, Sutter et al.~\cite{sutter2024dynamic}, and Elder et al.~\cite{elder2024survey} present them as proper SLRs.

Aydos et al.~\cite{aydos2022security} conduct a systematic mapping of web application security testing, screening 154 candidate articles down to 80 (2005–2020) through author voting against defined inclusion/exclusion criteria. They organize the field around four groups of research questions covering contribution and research-facet types, security-testing tools (tools used, license type, static-vs-dynamic analysis technique, automation level, false-positive/negative handling), security aspects (vulnerability type per OWASP Top 10, violation type, SDLC coverage), and the system under test (web application type, development language, connection type).

Alazmi et al.~\cite{alazmi2022systematic} conduct an SLR on web application vulnerability scanners (WVSs), evaluating 90 papers to catalog 30 scanners and their characteristics, of which only 12 had at least one quantitative effectiveness assessment (across 15 evaluation studies).
They find that scanner evaluations overwhelmingly test just two OWASP Top Ten categories, SQL injection (13/15 studies) and XSS (8/15), and that reported detection rates vary widely across evaluations, motivating calls for more comprehensive and consistent effectiveness benchmarking.

Ahsan et al.~\cite{ahsan2024systematic} conduct an SLR on search-based software security testing (i.e., the use of metaheuristics to generate security test cases against test-adequacy criteria) following Kitchenham's guidelines and analyzing 122 primary studies published between 2001 and 2022.
The review is structured around four research questions and classifies the vulnerability types tackled with metaheuristics (predominantly XSS, SQLi, program crashes, and XML injection), the metaheuristic algorithms used, and the fitness functions employed for code coverage and security testing, and identifies open challenges and future directions.

Sutter et al.~\cite{sutter2024dynamic} conduct an SLR on dynamic security analysis for Android, e.g., network monitoring, system-call tracing, dynamic taint analysis, reviewing 43 publications to establish a taxonomy and classification scheme for the field.
The review characterizes the published novel tools and techniques, their underlying research objectives (app security, privacy, malware detection, test automation), the prevalent constraints and shortcomings (code-coverage limits, analysis obstruction, app-selection adequacy, non-deterministic behavior), and prominent research trends.

Elder et al.~\cite{elder2024survey} survey software vulnerability exploitability assessment methods, i.e., techniques that estimate the likelihood or feasibility of exploiting a vulnerability, to help practitioners prioritize mitigation.
They selected 59 studies (disseminated through 72 papers) and four industry standards via active-learning-assisted search and snowballing.
They organize the field into three approach families: the original manual CVSS scoring, automated deterministic assessments (most commonly program-state-based), and automated probabilistic assessments (most commonly learning-based).

As a result, \textbf{none of the existing surveys focus on AEG and ASTG simultaneously}, highlighting the need for a better characterization.
In this respect, our study aims to provide a comprehensive, domain-agnostic SLR that covers both AEG and ASTG approaches.
Table~\ref{tbl:related-surveys} also reports the recurring dimensions analyzed in existing secondary studies.
Our survey does not restrict itself to a selection of vulnerabilities and software types; it analyzes the input and output of the resulting probe and assesses the automation level and tool availability.
This makes it the most comprehensive analysis existing so far that addresses both exploits and security test generation approaches.

\section{Survey Methodology}
\label{sec:methodology}
This survey follows the guidelines for conducting systematic reviews in software engineering recommended by Kitchenham et al.~\cite{keele2007guidelines,kitchenham2009systematic,kitchenham2013systematic}.
To this end, we first defined the objective of this review and formulated the research questions (Section~\ref{sec:research-questions}).
We then defined the search string, screening filters, and exclusion criteria to select relevant primary studies on automated exploit generation and security testing; we also performed backward snowballing on the primary set of studies to cover additional relevant studies that our search string may have missed (Section~\ref{sec:paper-collection}).
Finally, we reviewed the selected papers in depth and extracted key facts to address our research questions (see Section~\ref{sec:paper-organization}).

\subsection{Survey Goal and Research Questions}
\label{sec:research-questions}

This literature review aims to systematize the knowledge in \textbf{automated exploit generation (AEG) and security test generation (ASTG) targeting software vulnerabilities}.
As mentioned in Section~\ref{sec:intro}, we collectively referred to both types of approaches as \textbf{Automated Probe Generation (APG)}.
This review aims to map existing studies published in the field (and thus outline its scope) and to identify the proposed approaches, their underlying generation techniques, their peculiarities, and potential connections among them.

This review has a double \textit{purpose}.
On the one hand, it aims to attract the attention of software engineering and cybersecurity \textbf{researchers} to this research field and to foster the development of modern solutions.
On the other hand, it provides software \textbf{developers} with a catalog of tools they can reuse in their activities.
To achieve these objectives, we formulated six research questions (RQs).

APG can be applied to various software and vulnerability types.
For instance, approaches range from targeting binary programs launched via the command-line interface to targeting web applications that receive input via web APIs.
The type of software they target is inevitably related to the \textit{programming language} they can support, or whether they are language-agnostic, as different approaches leverage specific language characteristics.
For example, the C memory model differs from the \textsc{Java} memory model, enabling memory-related vulnerabilities.
Consequently, programming languages and software types are inevitably associated with the vulnerabilities that can be exploited or tested, such as stack-based buffer overflows (typical of C/C++ programs) or SQL Injection (typical of PHP or \textsc{Java} web applications).
We also checked whether the approaches target specific known vulnerabilities or can also discover zero-day vulnerabilities.

\newrqbox{1}{\rqOne}

Before starting this review, we conducted a preliminary inspection of publications in the field and observed that there is no shared definition of \textit{`exploits'} and \textit{`security tests'} among them.
For example, Liu et al. (\ref{S43}) defined the outputs generated by their approach to exploit SQL Injection vulnerabilities as \textit{`test cases'}; however, these outputs are actually malicious SQL snippets that should be classified as \textit{`exploits'}.
Based on the definitions provided in Section~\ref{sec:intro}, the differences lie in their goal and structure.
For this review, we did not select papers based on \textit{how the paper used the terminology}; instead, we assessed whether the generated artifacts aligned with either of the definitions we provided.
An additional objective of this review was also to clarify the blurry boundary between the two main types of probes (exploits and security tests).

Furthermore, in this preliminary analysis, we also observed that probes can have different shapes.
For instance, a generated exploit could simply be a specific string that triggers the vulnerability, as well as a runnable binary armed with shellcode.
Similarly, the security test case could range from a single input datum to a complete test case.

Besides, the mere execution of the generated probes might not immediately guarantee they work as intended.
Generated exploits, once executed, may provide no feedback on successful execution (e.g., no reverse shell is spawned), whereas generated test cases may lack automated assertions confirming the presence of an exploitable vulnerability.
Specifically, we are interested in the mechanism used to verify probe correctness, i.e., the verification oracle.

\newrqbox{2}{\rqTwo}

Studies introducing APG approaches often adopt known techniques to generate probes and reuse established tools, such as for source code analysis.
For instance, dynamic symbolic execution is a popular technique in this field that is used to find test inputs that execute a specific path in the target program, which may be necessary to trigger a vulnerable state.
The engine for symbolic execution is often not re-implemented from scratch but reused from frameworks such as \textsc{Klee}~\cite{cadar2008klee}.
Nevertheless, the studies may propose completely novel approaches.

\newrqbox{3}{\rqThree}

APG approaches require some data or information to generate the probe.
For example, white-box approaches require access to the source code of the targeted software, whereas black-box approaches require only the program's executable or an API endpoint.
Some approaches may require an initial PoC exploit that triggers the appropriate vulnerable state, or even text specifications (e.g., attack grammars) for the desired behavior of the exploit/test.

\newrqbox{4}{\rqFour}

APG approaches are often accompanied by experiments (of varying scope) that validate and demonstrate their effectiveness.
For instance, an approach could have been applied to a set of programs known to be vulnerable, and their success rates could have been measured. 
We mapped the evaluation datasets used, their nature (i.e., real or artificial software), size, and whether they are still publicly accessible.

\newrqbox{5}{\rqFive}

APG tools may differ in their levels of automation, requiring the user (e.g., the software developer) to provide a variable number of commands or actions.
We also mapped the list of tools proposed in the reviewed studies, the setup required, the maturity of their documentation, and their current accessibility.

\newrqbox{6}{\rqSix}

\subsection{Paper Collection and Selection}
\label{sec:paper-collection}

\subsubsection{Search String} To collect the studies for our systematic review, we designed the following search string:

\begin{tcolorbox}[title=Search String]
\textit{( automat* AND ( secur* OR vulnerab* ) AND ( test OR exploit ) AND generat* )}
\end{tcolorbox}

With this string, we sought to identify studies on \textbf{automated approaches that generate exploits or tests for software vulnerabilities}.
This string was refined through several preliminary experiments that tested alternative search terms and was selected based on the relevance of the top results and the overall number of hits.

We increase the chances of finding more relevant papers by using the asterisk operator in the search query.
For example, the search keyword \textit{``vulnerab*''} can retrieve papers containing the words \textit{``vulnerable''}, \textit{``vulnerability''} or \textit{``vulnerabilities''}.
We ran the query on each paper's title, abstract, and keywords, which is generally sufficient to capture the most relevant results.

We first performed our search on Scopus,\footnote{Scopus website: \url{https://scopus.com}} as it is widely recognized as the largest structured digital database, including studies from specialized venues such as USENIX Security and NDSS (Network and Distributed System Security Symposium).
As Scopus might miss some results, we included other established databases to ensure the coverage of different disciplines and publication types: ACM Digital Library,\footnote{ACM Digital Library website: \url{https://dl.acm.org/}} IEEE Xplore,\footnote{IEEE Xplore website: \url{https://ieeexplore.ieee.org/}} Web of Science,\footnote{Web of Science website: \url{https://clarivate.com/products/web-of-science/}} and DBLP\footnote{DBLP website: \url{https://dblp.org/}}.
The search retrieved an initial list of \textbf{3,309 papers} through December 31, 2024 (inclusive).

\subsubsection{Screening Filters}
We applied five \textbf{Screening Filters (SF1--SF5)} to exclude papers that were not relevant to our scope or were difficult to analyze, e.g., did not appear in high-ranked venues or due to language barriers.
Such filters are based on common criteria used in software engineering literature reviews~\cite{kitchenham2015evidence,kitchenham2013systematic}.
We discard papers that meet at least one of the following filters:
\begin{itemize}
	\item \textbf{SF1:} The paper is published before 2005 (i.e., we only retain papers published in the last two decades).
	\item \textbf{SF2:} The paper is not fully written in English or does not have an accessible text version.
	\item \textbf{SF3:} The paper has fewer than 6 pages.
	\item \textbf{SF4:} The paper is not peer-reviewed nor published in either a conference proceeding or a journal issue.
	\item \textbf{SF5:} The paper is a duplicate or a shortened version of another paper previously found (e.g., journal extensions are always preferred over the conference versions).
\end{itemize}

Consequently, \textbf{2,448 papers} (74\%) were retained for the next steps.

\subsubsection{Exclusion Criteria and Snowballing}
\label{subsub:exclusion}
The paper was subject to a number of exclusion criteria (EC); any paper that met any of them was discarded.
To select the optimal set of criteria, we first conducted a \textit{``calibration''} phase on a sample of 100 papers drawn from the 2,448 screened papers involving four co-authors of this work.
Specifically, we selected the 100 most cited papers, as recorded in Scopus at the time of the study. 

\begin{table}
\centering
\caption{Exclusion criteria for selecting papers during the relevance assessment process.}
\label{tbl:relevance-criteria-final}
\resizebox{\linewidth}{!}{%
	\begin{tabular}{llp{9cm}}
        \toprule
        \textbf{ID} & \textbf{Name}                           & \textbf{Description} \\
        \midrule
        \textbf{EC1} 		& Out-of-Scope & The paper is not related to Computer Science and Engineering. \\
        \textbf{EC2} 		& Secondary-Study & The paper is a survey, review, or a Systematization of Knowledge (SoK). \\
        \textbf{EC3}          & No-Generation & The paper does not describe an approach that generates exploits or security tests for security vulnerabilities. Papers that would fall under this criterion are those: (1) not focused on security vulnerabilities, (2) making only an evaluation, benchmarking, or comparison of existing approaches, (3) introducing tools, platforms, or wrappers of existing approaches without conceptual additions, and (4) transforming the format of existing exploits/tests without changing their behaviors. \\
        \textbf{EC4} 		& Fuzzing & The paper's main focus is on developing a new fuzzing approach. However, papers that employ existing fuzzing tools as part of the exploit or security test generation workflow \textit{do not} fall under this criterion (i.e., they are \textit{not excluded}). \\
        \textbf{EC5}		& Not-High-Quality-Venue & The paper is not published in A/A* conferences (CORE\tablefootnote{Conference Rankings: \url{https://portal.core.edu.au/conf-ranks/}}) nor in a Q1 Journals in Computer Science (SCIMAGO\tablefootnote{Journal Rankings: \url{https://www.scimagojr.com/journalrank.php?area=1700}}). \\   
        \midrule
        \textbf{EC6} 		& No-Software-Vulnerabilities & The paper presents an approach that does not target vulnerabilities specifically in software applications. Papers targeting vulnerabilities in network and system configuration (e.g., misconfiguration of firewalls), or in hardware systems (e.g., IoT devices), meet this criterion (i.e., they are excluded). \\
        \textbf{EC7} 		& Target-Specifications & The paper presents an approach that targets specifications (e.g., protocols, formal models, policies, textual requirements). \\
        \bottomrule 
    \end{tabular}
}
\end{table}

As described in Table~\ref{tbl:relevance-criteria-final}, we formulated an initial set of five exclusion criteria, which we extended to seven during calibration.
The initial criteria EC1, EC2, and EC3 reflect the main focus of this review: primary studies on automated probe (exploit and security test) generation.
The initial criterion EC4 was included in our initial set of criteria for scoping reasons, as literature reviews on fuzzing already exist~\cite{manes2019art, mallissery2023demystify, zhu2022fuzzing} and we did not want to overlap with such work, as it would not add value.
In fact, there is a large body of work on fuzzing
that has overshadowed other types of approaches, which have remained outside review studies so far.
Finally, the initial criterion EC5 was introduced during calibration to focus our work on very high-impact works and to make the size of the literature review more manageable---as often done in other literature reviews.

During calibration, each co-author was instructed to independently read the titles, abstracts, and keywords of 100 papers to determine whether they met a specified criterion.
In addition, if an inspector believed a paper should be discarded for reasons not covered by the initial criteria, they could express their opinion in text.
At the end of the independent inspection, we consolidated the results in a joint meeting.
We first reviewed all cases of conflict, i.e., an inspector wanted to discard a paper that another author wanted to preserve, and we jointly discussed them until a consensus was reached.
We then reviewed the proposed missing exclusion criteria collectively and discussed their suitability.
This resulted in two additional exclusion criteria (EC6 and EC7), which were introduced to improve the alignment of the results to the scope of our review, which specifically target vulnerabilities in software applications that could be identified in code- or design-level artifacts.

The calibration excluded 91 papers (i.e., 9 survived). With a stable set of exclusion criteria, we divided the remaining 2,348 papers 
into four batches and assigned each batch to one of the four co-authors involved in the calibration phase.
Here too, the papers were inspected through their titles, abstracts, and keywords.
This phase led to the exclusion of 2,295 papers; considering the 9 papers retained from the calibration batch, we ended up with \textit{53 retained papers} (2\% of the total).

Lastly, we performed \textit{backward snowballing} on the set of 53 papers, i.e., by navigating their accessible references, we collected a total of 1,839 papers.
We removed duplicates already processed in the initial search results, re-applied the screening filters, and discarded papers meeting our exclusion criteria, yielding \textit{18 additional papers}.
We ended up with a total of \textbf{\totalPapers papers}, which we approved for the review phase.

\subsection{Paper Review and Categorization}
\label{sec:paper-organization}

\begin{figure}[t]
	\includegraphics[width=\textwidth]{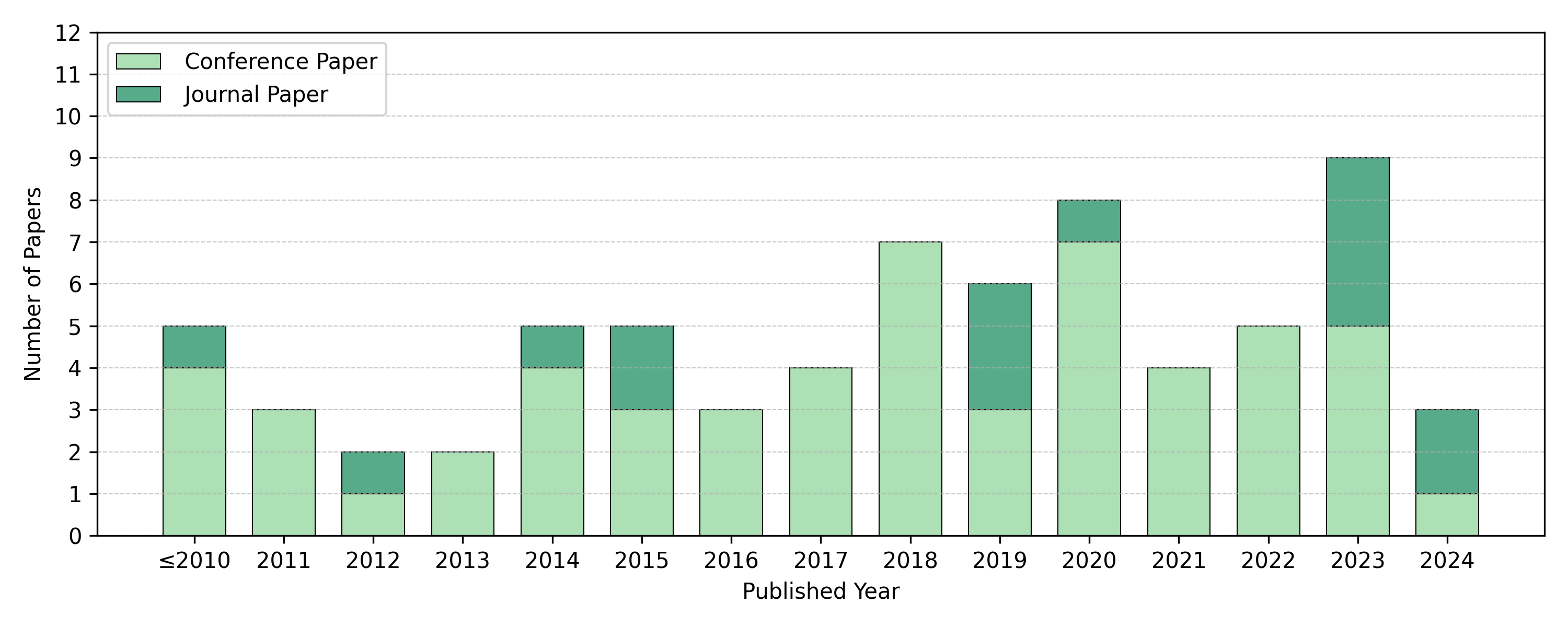}
	\caption{Number of published papers per year with publication types.}
	\label{fig:papers-by-year}
\end{figure}

\begin{figure}[t]
	\includegraphics[width=0.5\textwidth]{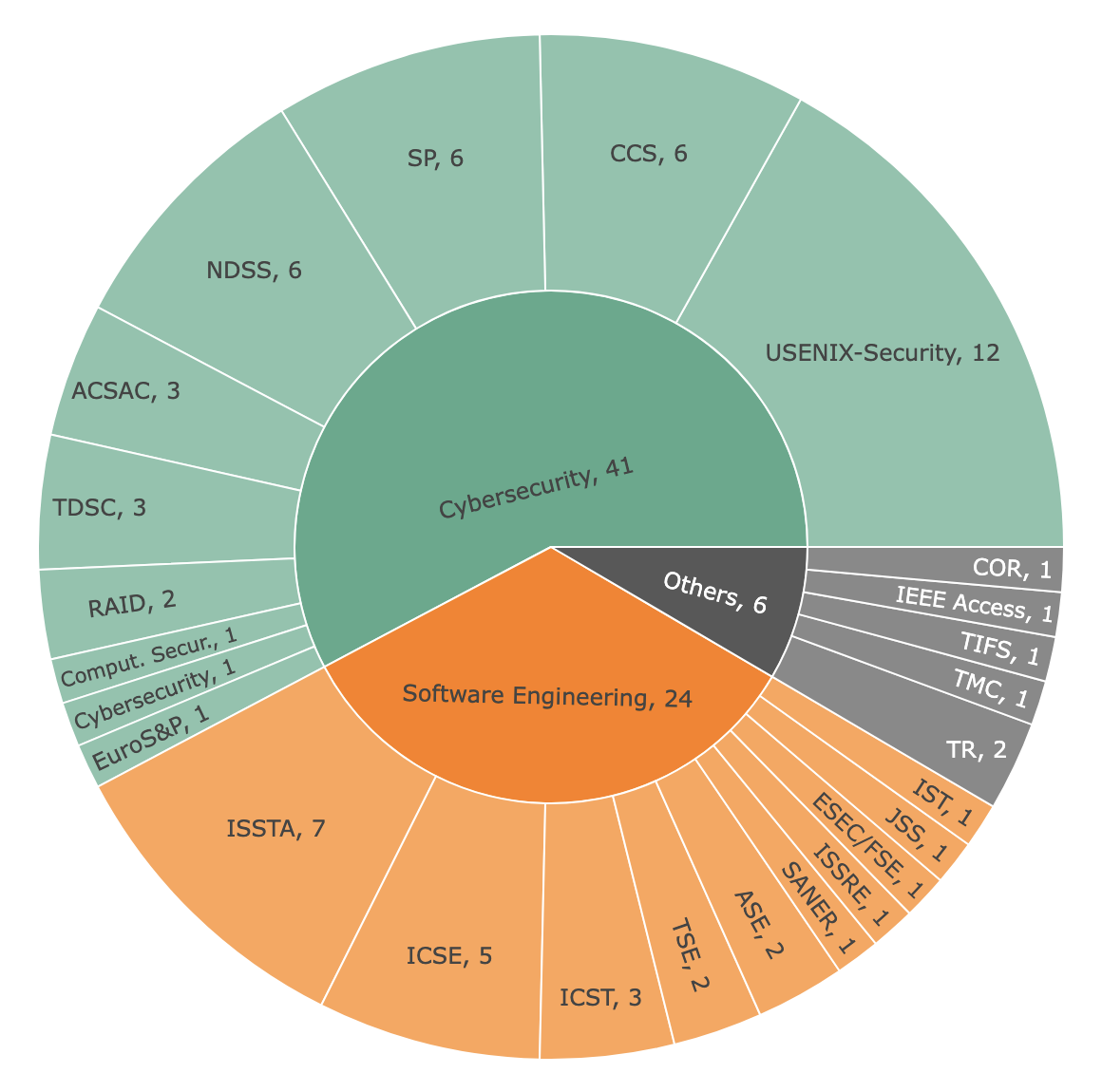}
	\caption{Number of published papers by venue.}
	\label{fig:papers-by-venue}
\end{figure}

We distributed the \totalPapers papers in batches among four co-authors and processed them to collect relevant facts to answer our research questions.
We release the review form and the raw data produced by the reviewers in our supplementary materials~\cite{replicationpackage}.

After the review, we grouped papers that belonged to the same \textit{family of studies}, e.g., when one paper is a follow-up or evolution of essentially the same approach.
In total, we identified \textbf{\totalStudies studies}.
The full list of studies reviewed is reported in Appendix~\ref{appx:papers}, along with key metadata for each study.
Throughout the review, \textbf{we refer to the studies rather than individual papers.}

Figure~\ref{fig:papers-by-year} shows the papers categorized based on the publication type (conference vs. journal) and publication year.
Most were published in the latter decade (2014--2024), totaling 58 out of \totalPapers (82\%) papers.
Figure~\ref{fig:papers-by-venue} shows the distribution of papers across research venues. A total of 55 out of \totalPapers (77\%) papers were presented in conference proceedings, primarily in software engineering (e.g., ICSE, ESEC/FSE, ASE) and cybersecurity (e.g., CCS, USENIX Security, S\&P, RAID). The remaining papers appeared in journals, including high-impact venues such as IEEE Transactions on Software Engineering (TSE), IEEE Transactions on Dependable and Secure Computing (TDSC), and IEEE Transactions on Reliability (TR).

\section{Review Results}
\label{sec:results}

\subsection{RQ$_1$ -- Types of Software \& Vulnerabilities}

\textcolor{black!75}{\textit{``\rqOne''}}

In this research question, we first analyzed the types of software and vulnerabilities that APG approaches target (i.e., the asset being executed by the generated probes).
We categorized the results \textbf{primarily by software type}, as specific vulnerabilities are typically inherent to certain software domains. For example, Cross-Site Scripting (XSS) and SQL Injection (SQLi) are predominantly found in web-facing software (e.g., web applications).
Table~\ref{tbl:rq1-results-old} categorizes the studies by targeted software type and reports the related programming languages and vulnerability types.

\begin{table}[t]
	\centering
	\footnotesize
	\caption{Targets of the APG studies. The perspective is that of the targeted software, which may be implemented in different programming languages and may be affected by different vulnerability types. Consequently, a given programming language and vulnerability type may occur across multiple software types.}
	\label{tbl:rq1-results-old}
    \resizebox{\linewidth}{!}{%
	\begin{tabular}{p{2.7cm}p{2.5cm}p{5.5cm}p{3cm}}
    \toprule
		\textbf{Targeted Software} & \textbf{Targeted PLs} & \textbf{Targeted Vulnerability Types} & \textbf{Studies} \\
		\midrule
		Web-facing Software & Java, Python, PHP, WXML & XSS, SQL Injection, CSRF, Resource Exhaustion, Mass Assignment, OSGI Injection, PHP Object Injection, Remote Code Execution, Access Control, SSO Misuse, REST API Violations, Unrestricted File upload, Server-Side Template Injection (SSTI) & \ref{S04}, \ref{S08}, \ref{S10}, \ref{S12}, \ref{S14}--\ref{S17}, \ref{S21}--\ref{S24}, \ref{S42}--\ref{S43}, \ref{S45}--\ref{S46}, \ref{S52}--\ref{S53}, \ref{S56}, \ref{S60}, \ref{S62}--\ref{S63}, \ref{S66}--\ref{S67} (24 total) \\
        \hline
		Command-line Utilities & C/C++, Binary-based, Assembly, Python & Memory-based, Sensitive Information Leakage, Out-of-Bounds, Logic Errors & \ref{S01}, \ref{S03}, \ref{S06}--\ref{S07}, \ref{S09}, \ref{S18}--\ref{S19}, \ref{S31}, \ref{S37}, \ref{S39}, \ref{S44}, \ref{S48}, \ref{S55}, \ref{S58}--\ref{S59}, \ref{S64}--\ref{S65} (18 total)\\
        \hline
		End-user Applications & C/C++, Binary-based, Python & Memory-based, Input Validation, Integer overflow & \ref{S02}, \ref{S13}, \ref{S20}, \ref{S29}, \ref{S35}, \ref{S47}, \ref{S49} (7 total) \\
        \hline
		Operating Systems & C/C++, Binary-based & Memory-based (UAF, Heap), Unix file-system race, Out-of-Bounds & \ref{S05}, \ref{S11}, \ref{S32}, \ref{S36}, \ref{S38}, \ref{S40}, \ref{S54} (7 total) \\
        \hline
		Android Platform & Java, Java Bytecode, JavaScript & Android ICC, XSS, Inconsistent Security Policy Enforcement (IPSE), Task Hijacking, Access Control, Intent Abuse, Uncaught Exception & \ref{S25}--\ref{S28}, \ref{S30}, \ref{S41} (6 total)\\
        \hline
		Regex Matchers & Agnostic & Regular Expression Denial of Service & \ref{S34}, \ref{S50}--\ref{S51}, \ref{S61} (4 total)\\
        \hline
		Smart Contracts & EVM Bytecode, C/C++, WASM & Memory-based, Self-Destructible, Call Injection & \ref{S33}, \ref{S57} (2 total) \\
		\bottomrule
	\end{tabular}
    }
\end{table}

\subsubsection{The Software Type Perspective}

Across the \totalStudies primary studies, we identified \textbf{seven recurring software types}.
The first two, i.e., \textit{web-facing software} and \textit{command-line utilities} are the most targeted software domains, accounting for approximately 63\% (i.e., 42 out of 67) of current APG studies.
Other software categories that have received comparatively less attention include \textit{end-user applications} (10\%), \textit{operating systems} (10\%),  and the \textit{Android platform} (9\%). The least studied domains are \textit{regular expression matchers} (6\%) and \textit{smart contracts} (3\%).

\paragraph{Web-facing Software}
Web-facing software refers to any application or service hosted on a server and accessible to users via HTTP-based interfaces (e.g., a web browser), typically served over the internet.
This group also includes web frameworks that allow the simple creation of other websites, i.e., content management systems (CMS).
A total of 24 studies (36\%) focused on web-facing software, often opting for popular applications like \texttt{HotCRP}, \texttt{WordPress}, and \texttt{Magento}.
Nearly half of them (10 out of 24) generate security tests, adopting techniques like black-box (\ref{S24}) or mutation-based testing (\ref{S15}, \ref{S23})---more detail on these categories is given in Section~\ref{subsec:model-based-generation}.
The remaining studies (14 out of 24) introduce approaches specifically designed to generate malicious web input payloads and exploits. For example, Liu et al. (\ref{S43}) presented \textsc{DeepSQLi}, adopting a deep learning-based neural language model to transform benign SQL user inputs (e.g., \texttt{user\_id=`2'}) into malicious inputs (e.g., \texttt{user\_id=`2;delete from users'}) that constitute an SQLi exploit.

In this category, PHP, Java, and Python are the most targeted programming languages as they are widely used to build web systems.
Besides, we also found studies targeting specific API styles adopted by the web applications, such as SOAP (\ref{S15}, \ref{S23}) and RESTful (\ref{S45}, \ref{S60}).
Regarding vulnerability types, injection vulnerabilities are the most frequent targets in web-based software, spanning XSS (CWE-79), SQL Injection (CWE-89), CSRF (CWE-352), OSGI Injection (CWE-94), PHP Object Injection (CWE-502), and Server-side Template Injection (SSTI) (CWE-1336).
Additionally, the reviewed studies also targeted uncommon vulnerability types that are only applicable to web applications having a specific design, such as ``SSO Misuse'' (CWE-287) affecting applications employing the Single Sign-On (SSO) authentication scheme, or ``Mass Assignment'' (CWE-915) and ``REST API Violations'' affecting RESTful applications.

\paragraph{Command-line Utilities}
Command-line utilities include programs designed to perform specific tasks, e.g., network operations or file processing, by accepting text-based commands and options from a terminal, as well as shared libraries that expose functionality for other programs.
12 studies (18\%) in our review proposed approaches for generating probes for command-line programs (e.g., \texttt{grep}, \texttt{gzip}, \texttt{cvs}, \texttt{coreutils}, \texttt{binutils}) or shared libraries (e.g., \texttt{libjpeg}, \texttt{libtiff}) widely used in Unix-like systems.
These programs and libraries are mostly written in C/C++, Assembly, or, occasionally, Python.
Therefore, memory-based vulnerabilities such as buffer overflows (CWE-119) are the most frequent targets in this software type.
Furthermore, there are approaches that focus on Out-of-Bounds Read (CWE-125) and Write (CWE-787) vulnerabilities (\ref{S44}), Logic Errors (CWE-840) (\ref{S44}), as well as Sensitive Information Leakage (CWE-200) (\ref{S18}).

\paragraph{End-user Applications}
End-user applications are comprehensive software programs that are used through a Graphical User Interface (GUI) by non-technical users to perform everyday tasks, such as media players (\texttt{GSPlayer}) and web browsers (\texttt{Chrome}).
In total, seven studies (11\%) targeted end-user applications runnable on Unix-like systems or Windows.
The most common applications belonging to this category are \texttt{FileZilla FTP Server}, \texttt{Chrome}, and \texttt{LibTorrent}.
These software applications are provided to APG approaches either as source code (primarily C/C++) or as binaries.
As with \textit{Command-line utilities}, memory-based vulnerabilities are the most common targets in desktop programs. Moreover, Input Validation (CWE-20) (\ref{S02}) and Use of Uninitialized Variable (CWE-457) (\ref{S13}) are also addressed.

\paragraph{Operating Systems}
This group includes operating system images and kernel versions.
Seven studies (11\%) targeted operating systems, primarily Linux kernel versions such as \texttt{Linux kernel 4.x}, often to discover unknown vulnerabilities.
The majority of the targeted vulnerabilities are memory bugs, including Use-After-Free (CWE-416) (\ref{S32}, \ref{S38}), Heap-based buffer overflows (CWE-122) (\ref{S54}), and Out-of-Band (CWE-125, CWE-787) (\ref{S40}).
Some studies targeted four different kinds of Unix-like systems, such as Cai et al. (\ref{S05}), who attempted to exploit Unix file-system race vulnerabilities in Linux, FreeBSD, OpenSolaris, and OpenBSD.
On the other hand, we found studies that targeted custom operating systems, such as the study by Cotroneo et al. (\ref{S11}), in which the approach generated robustness tests.

\paragraph{Android Platform}
This category includes both Android applications and the Android ecosystem itself, i.e., the operating system, runtime, and services.
This is the only platform for mobile software targeted by APG approches, with six studies (9\%).
The most frequently exploited vulnerabilities are related to Android Inter-Component Communication (CWE-926, CWE-927) (\ref{S26}, \ref{S27}), which are highly Android-specific.
However, the study by Wu et al. (\ref{S25}) targeted uncaught exceptions in Android apps by generating proof-of-concept exploits via fuzzing.
We also found special cases, such as the study by Yang et al. (\ref{S28}), which attempted to exploit XSS (CWE-79) vulnerabilities in Android hybrid applications implemented with web technologies.

\paragraph{Regex Matchers}
Regex matchers refer to programs or libraries, or even just parts of them, that use regular expression patterns to search, validate, or manipulate strings.
Four studies (\ref{S34}, \ref{S50}--\ref{S51}, \ref{S61}) tested whether regular expression matchers could lead to an infinite loop, causing a Regular Expression Denial of Service (CWE-1333).
This occurs by crafting incorrect input (e.g., very long or with specific patterns) that causes the regex matcher to perform backtracking exponentially.
This is often caused not by a bug in the regex matcher itself (which is often from an established library), but by the regular expression pattern, which causes the regex matcher to explore a huge number of possible paths.
This is often due to nested quantifiers, overlapping alternations, and ambiguous repetitions.
For this reason, these approaches are language-agnostic and independent of the software system in which they are used, as long as the system uses regex-based matching.

\paragraph{Smart Contracts}
Smart contracts are self-executing programs stored on a blockchain that automatically enforce the terms of an agreement when predefined conditions are met.
They are typically written in high-level languages such as Solidity or Rust, which are then compiled into Ethereum Virtual Machine (EVM) bytecode or WebAssembly (WASM) for execution.
In the two studies (\ref{S33}, \ref{S57}), the authors introduced new types of vulnerabilities in smart contracts (i.e., memory-related, self-destructible, and call injection) and proposed approaches to exploit them.

\subsubsection{The Programming Language Perspective}

From the programming language perspective, the reviewed studies show a strong focus on C/C++ or binary-based software, with 27 out of 67 studies (40\%).
This reflects the prominence of memory-related vulnerabilities in command-line utilities, end-user applications, and operating systems.
Java was targeted in 15 studies (22\%), while PHP appeared in 13 studies (19\%), both predominantly in the context of web-facing software.
By contrast, Python and JavaScript are targeted by only three (4\%) and four studies (6\%), respectively, despite their widespread adoption in modern software development.
Lastly, 9 studies (13\%) are language-agnostic or operate on artifacts such as regular expressions.
This distribution suggests that language support is largely determined by the targeted vulnerability class and by the availability of mature analysis tools.

\subsubsection{The Vulnerability Type Perspective}

From a vulnerability-type perspective, the reviewed studies primarily focused on a restricted set of well-established vulnerability types.
Memory-related vulnerabilities are the most frequent target, appearing in 28 out of 67 studies (42\%).
This group includes buffer overflows, heap-related vulnerabilities, use-after-free bugs, ROP-oriented exploits, and other binary-level weaknesses.
Web- and injection-related vulnerabilities constitute the second-largest group, appearing in 18 studies (27\%); these include SQL Injection, XSS, XML Injection, PHP Object Injection, Command Injection, File Inclusion, and Server-side Template Injection.
Access Control, Policy, and API-related issues appear in 7 studies (10\%), while denial-of-service vulnerabilities are less common with four studies (6\%) targeting ReDoS issues.
Smart-contract-specific vulnerabilities appear in only two studies (3\%).
Overall, current APG research prioritizes vulnerabilities that are easier to model and automatically detect at the code level, while more complex vulnerabilities that depend on the overall software design remain comparatively underexplored.
We note that we did not explicitly map each study to specific CWEs, as most did not mention them and instead referred to broader vulnerability types.

\subsubsection{Targeted Vulnerabilities: Known vs. Zero-day}

We investigated whether APG approaches target known or zero-day vulnerabilities.
We classified an approach as \textit{targeting known vulnerabilities} when it requires specific information about the vulnerability to start the generation process that is commonly available after discovery/disclosure, e.g., the patch, the vulnerability location, a proof-of-concept input, or a crash-triggering input.
Conversely, we classified approaches as \textit{targeting zero-day vulnerabilities} when they can analyze the target without prior knowledge of whether a vulnerability even exists, even when they assume a known attack class or security property.

Overall, 31 out of 71 studies (44\%) target known vulnerabilities, while 40 studies (56\%) target zero-day vulnerabilities.
Under this coding, the two categories are mutually exclusive, as we classified the studies based on the input required by the generation approach.
In this respect, we did not classify according to the subjects involved in the evaluation set.
Therefore, evaluating an approach on CVEs, Capture the Flag (CTF) challenges, benchmarks, or deliberately vulnerable applications does not necessarily imply that it targets only known vulnerabilities.

The near-even distribution highlights two complementary uses of APG approaches.
Approaches targeting known vulnerabilities support (directly or indirectly) \textit{reproduction} and \textit{risk assessment} by leveraging vulnerability-specific pieces of information.
In contrast, zero-day approaches support \textit{vulnerability discovery} and \textit{triage} without requiring a pre-identified flaw.
However, the latter should not be interpreted as unconstrained vulnerability discovery: they typically search for a predefined attack class, security property, or violation condition.

\rqanswer{1}{
APG studies target seven main software types, led by web-facing software (36\%) and command-line utilities (18\%). Target vulnerabilities are typically associated with the software domain, e.g., studies on web applications focus on injection bugs (XSS, SQLi), whereas studies on low-level software (command-line, OS, end-user apps) focus on memory flaws. Target software is mostly written in C/C++ or binary-based (40\%) for memory vulnerabilities, and PHP (19\%) or Java (22\%) for web software. Finally, 56\% of studies can probe zero-day vulnerabilities, though 44\% still rely on prior vulnerability-specific pieces of information.
}

\subsection{RQ$_2$ -- Generated Artifacts}
\label{subsec:rq2}

\textcolor{black!75}{\textit{``\rqTwo''}}

In this research question, we analyzed the types of artifacts (i.e., the probes) returned by APG approaches.
We first analyzed the nature of the probe itself, i.e., the kind of exploit or test case.
Then, we analyzed the type of \textbf{verification oracle} that is produced alongside or directly embedded in the probe.
This oracle refers to the mechanism used to confirm whether the probe successfully exploited or tested the target vulnerability.
Table~\ref{tbl:rq2-output} summarizes the generated outputs and verification oracles across the \totalStudies studies.

Overall, the \totalStudies studies could be divided into those that generated exploits (or part of it) and those that generated test cases (or part of it).
We determined whether a probe adhered to the definition of exploits or test cases (as in Section~\ref{sec:intro}) based on their \textit{structure and goal}, rather than on what the study called the generated artifact.
Namely, we identified 51 studies (76\%) that generated exploits and 16 (24\%) that generated security test cases.
Thus, exploits appear to be the largely preferred type of probe being targeted by academic research.

In several cases (44 studies, 66\%) the generated probe was \textit{incomplete}, i.e., the APG approach only generated part of it.
This often consists of just the \textit{payload}, i.e., the data or code that delivers the concrete or harmless attack (as defined in Section~\ref{sec:intro}).
Specifically, 35 of 51 studies (69\%) on exploits produced an incomplete probe, while among studies on test cases, 9 of 16 studies (56\%) did.
Thus, most approaches are able to produce only elements of a full probe, suggesting the difficulty of effective end-to-end probe generation.

Notably, \textsc{Tunter} (\ref{S59}) can generate either an exploit payload or a proper exploit that carries that payload.
Namely, it first identifies an exploitable program state and generates a concrete input that reaches it.
This first result is the payload of an exploit.
Then, it tries to satisfy the additional constraints needed to turn the input into a complete exploit, producing a second result.
In its evaluation, \textsc{Tunter} generated the complete exploit in only 11 of 16 cases; in 5 cases, it produced only a payload.
We considered this approach as part of both classes.

As the studies concerning exploits were the majority (76\% of the total), we also provided a second-level classification for better characterization.
Among the 17 approaches that can generate full exploits, we identified two subtypes:
\begin{itemize}
    \item \textit{Runnable Program} (13 studies), which is a self-contained executable artifact (program, script, malicious application, smart contract) that can deliver a complete attack (either harmful or harmless).
    \item \textit{Attack Sequence} (4 studies), which indicates an ordered list of application-level interactions with crafted input that together deliver an attack (either harmful or harmless). Each step can be an HTTP request, browser action, a blockchain transaction, etc. Later actions build on the results produced by earlier actions, e.g., a session, token, or object.
\end{itemize}

\begin{table*}[t]
\centering
\caption{Type of probes generated by the approaches introduced in the \totalStudies surveyed studies, including the verification oracles produced alongside.}
\label{tbl:rq2-output}
\resizebox{\linewidth}{!}{%
\begin{tabular}{c|l|l|p{3.2cm}|p{2.6cm}||c|l|l|p{3.2cm}|p{2.6cm}}
\hline
\textbf{Study} &
\textbf{Probe Type} &
\textbf{Partial/Full} &
\textbf{Subtype} &
\textbf{Oracle} &
\textbf{Study} &
\textbf{Probe Type} &
\textbf{Partial/Full} &
\textbf{Subtype} &
\textbf{Oracle}
\\
\hline
\ref{S01} & Exploit & Partial & Exploit Input & Output Analysis &
\ref{S35} & Exploit & Partial & Exploit Building Block & State Analysis \\
\ref{S02} & Exploit & Partial & Exploit Input & Manual Verification &
\ref{S36} & Exploit & Partial & Exploit Building Block & State Analysis \\
\ref{S03} & Test & Partial & Test Input Data & State Analysis &
\ref{S37} & Exploit & Partial & Exploit Building Block & State Analysis \\
\ref{S04} & Exploit & Partial & Exploit Input & Relational Analysis &
\ref{S38} & Exploit & Partial & Exploit Building Block & State Analysis \\
\ref{S05} & Exploit & Partial & Exploit Input & Output Analysis &
\ref{S39} & Exploit & Full & Runnable Program & Manual Verification \\
\ref{S06} & Exploit & Partial & Exploit Input & Observable Impact &
\ref{S40} & Exploit & Full & Runnable Program & State Analysis \\
\ref{S07} & Exploit & Partial & Exploit Input & State Analysis &
\ref{S41} & Exploit & Full & Runnable Program & Manual Verification \\
\ref{S08} & Exploit & Partial & Exploit Message & State Analysis &
\ref{S42} & Test & Full & - & State Analysis \\
\ref{S09} & Exploit & Full & Runnable Program & Observable Impact &
\ref{S43} & Exploit & Partial & Exploit Input & Output Analysis \\
\ref{S10} & Test & Full & - & Relational Analysis &
\ref{S44} & Exploit & Partial & Exploit Input & State Analysis \\
\ref{S11} & Test & Partial & Assertion-less Test Case & Observable Impact &
\ref{S45} & Exploit & Full & Attack Sequence & Output Analysis \\
\ref{S12} & Exploit & Partial & Exploit Message & Output Analysis &
\ref{S46} & Exploit & Partial & Exploit Message & Output Analysis \\
\ref{S13} & Exploit & Partial & Exploit Input & Observable Impact &
\ref{S47} & Exploit & Partial & Exploit Message & Observable Impact \\
\ref{S14} & Exploit & Partial & Exploit Input & State Analysis &
\ref{S48} & Test & Full & - & State Analysis \\
\ref{S15} & Test & Partial & Test Input Data & State Analysis &
\ref{S49} & Exploit & Full & Runnable Program & Reference Match \\
\ref{S16} & Exploit & Partial & Exploit Building Block & Output Analysis &
\ref{S50} & Exploit & Partial & Exploit Input & State Analysis \\
\ref{S17} & Exploit & Full & Attack Sequence & Output Analysis &
\ref{S51} & Exploit & Partial & Exploit Input & State Analysis \\
\ref{S18} & Exploit & Partial & Exploit Input & State Analysis &
\ref{S52} & Exploit & Partial & Exploit Message & State Analysis \\
\ref{S19} & Test & Partial & Test Input Data & State Analysis &
\ref{S53} & Test & Full & - & State Analysis \\
\ref{S20} & Exploit & Partial & Exploit Input & Manual Verification &
\ref{S54} & Exploit & Partial & Exploit Input & State Analysis \\
\ref{S21} & Exploit & Partial & Exploit Input & State Analysis &
\ref{S55} & Exploit & Partial & Exploit Building Block & Reference Match \\
\ref{S22} & Exploit & Full & Attack Sequence & State Analysis &
\ref{S56} & Exploit & Partial & Exploit Input & Output Analysis \\
\ref{S23} & Test & Partial & Test Input Data & Output Analysis &
\ref{S57} & Exploit & Full & Runnable Program & State Analysis \\
\ref{S24} & Test & Full & - & Output Analysis &
\ref{S58} & Exploit & Full & Runnable Program & Reference Match \\
\ref{S25} & Exploit & Full & Runnable Program & Observable Impact &
\ref{S59}$^*$ & Exploit & Partial, Full & Exploit Primitive, Runnable Program & State Analysis \\
\ref{S26} & Exploit & Partial & Exploit Message & State Analysis &
\ref{S60} & Test & Partial & Assertion-less Test Case & State Analysis \\
\ref{S27} & Test & Partial & Assertion-less Test Case & State Analysis &
\ref{S61} & Exploit & Partial & Exploit Input & State Analysis \\
\ref{S28} & Test & Partial & Assertion-less Test Case & State Analysis &
\ref{S62} & Test & Partial & Assertion-less Test Case & Relational Analysis \\
\ref{S29} & Exploit & Full & Runnable Program & State Analysis &
\ref{S63} & Test & Full & - & Relational Analysis \\
\ref{S30} & Exploit & Full & Runnable Program & State Analysis &
\ref{S64} & Exploit & Partial & Exploit Building Block & Relational Analysis \\
\ref{S31}$^\dagger$ & Exploit & Full & Runnable Program & State Analysis, Observable Impact &
\ref{S65} & Exploit & Partial & Exploit Input & Observable Impact \\
\ref{S32} & Exploit & Full & Runnable Program & State Analysis &
\ref{S66} & Exploit & Partial & Exploit Message & Output Analysis \\
\ref{S33} & Exploit & Full & Attack Sequence & State Analysis &
\ref{S67} & Test & Full & - & Output Analysis \\
\ref{S34} & Exploit & Partial & Exploit Input & State Analysis & & & & &
\\ \hline
\multicolumn{10}{l}{
\textbf{Probe Type Total:}
Exploit: \textbf{51};
Test Case: \textbf{16}.
} \\ \hline
\multicolumn{10}{l}{
\textbf{Probe Completeness Total:}
Full: \textbf{23};
Partial: \textbf{44}.
} \\ \hline
\multicolumn{10}{l}{
\textbf{Subtype Total:}
} \\
\multicolumn{10}{l}{
\textit{Full Exploit}: Runnable Program: \textbf{13}; Attack Sequence: \textbf{4}.
} \\
\multicolumn{10}{l}{
\textit{Partial Exploit}: Input: \textbf{20}; Message: \textbf{7}; Building Block: \textbf{7}.
} \\
\multicolumn{10}{l}{
\textit{Partial Test Cases}: Assertion-less: \textbf{5}; Input Data: \textbf{4}.
} \\ \hline
\multicolumn{10}{l}{\textbf{Oracle Total:}
State Analysis: \textbf{35};
Output Analysis: \textbf{13};
Observable Impact: \textbf{8};
Relational Analysis: \textbf{5};
Manual Verification: \textbf{4};
Reference Match: \textbf{3}.
} \\ \hline
\multicolumn{10}{l}{
$^*$: \ref{S59} generates both a partial and a full exploit and is counted in both corresponding categories.
} \\
\multicolumn{10}{l}{
$^\dagger$: \ref{S31} verified the generated exploit in two ways, i.e., checking the heap layout and whether a shell was spawned; thus, it is counted in both categories.
} \\
\end{tabular}
}
\end{table*}

On the other hand, among the 34 approaches that can generate partial exploits, we identified three subtypes:
\begin{itemize}
    \item \textit{Exploit Input} (20 studies), which indicates the malicious data supplied through an accepted channel, e.g., a parameter value, a filename, a byte string, a regular expression, or a fuzzer-generated input. This artifact does not tell how to invoke the target, and it must be injected into an existing request, command, or invocation.
    \item \textit{Exploit Message} (7 studies), which indicates a complete message or invocation that the target can process and includes the malicious input, e.g., an HTTP request, a URL with specific query parameters, an Android intent, or a serialized object.
    \item \textit{Exploit Building Block} (7 studies), which indicates a low-level component used to construct an exploit by controlling a program's execution flow or memory state. This includes ROP chains, gadget chains, data-oriented programming sequences, heap-manipulation primitives, or shellcodes. This alone cannot function independently; it requires a triggered vulnerability (made by some prior exploit logic) to become effective.
\end{itemize}

Among the studies that generated security test cases, 7 could return complete test cases, which have all the data and logic to be launched against the target and also embed an automated oracle (in the form of \textit{assertions}) to verify whether the test succeeded or failed. These tests anyway require a specific \textit{test framework} to be invoked.
Instead, among the 9 studies that produced partial tests, we identified two subtypes:
\begin{itemize}
    \item \textit{Assertion-less Test Case} (5 studies), which indicates a test with the right vulnerability-triggering data and logic but lacking the assertions that verify whether it succeeded or failed; this is delegated to the tester or some external mechanisms that are not embedded in the test.
    \item \textit{Test Input Data} (4 studies), which indicates any data or information that is meant to be included in an externally provided test case, such as an SQL statement, an XML message, an HTTP request, or a template input. In other words, this is the payload of a test case.
\end{itemize}

Next, we analyzed the oracles that are produced by the APG approaches to verify that the generated probe achieves its intended goal.
We observed several distinct oracle types:
\begin{itemize}
    \item \textit{State Analysis} (35 studies, 52\%), which checks whether the execution reaches a vulnerable state, such as a sensitive sink, a target heap layout, a privilege-escalation condition, or another specified exploitation state.
    \item \textit{Output Analysis} (13 studies, 19\%), which evaluates the content of execution outputs, such as HTTP responses, generated messages, timing behavior, or other examinable results.
    \item \textit{Observable Impact} (8 studies, 12\%), which checks whether the probe produced a concrete side effect that impacts the target, such as spawning a shell (typical when the probe includes shellcode) or crashing the target (a form of denial of service).
    \item \textit{Relational Analysis} (5 studies, 7\%), which compares the result of executing the generated probe with a baseline execution (e.g., a safe one) or with an expected behavior. The vulnerability is exploited or reproduced if a certain expected difference is observed.
    \item \textit{Manual Verification} (4 studies, 6\%), which indicates that the study suggests that no automated oracle verification is performed in the APG approach; any assessment is left to the user's judgment.
    \item \textit{Reference Match} (3 studies, 4\%), which compares the generated artifact with a predefined reference output that acts as a ground truth. The comparison is typically done using exact string match or a similarity measure like BLEU.
\end{itemize}

We observed that, in most cases, probes are verified by checking the resulting program state after its execution (\textit{State Analysis} method).
This method is consistently adopted for both exploits and test cases, keeping the same global distribution.
Only the \textit{Observable Impact} verification method is mainly adopted by approaches generating exploits (namely, in 7 out of 8 cases).
This was expected, as this group includes shell spawning and program crashes, which are typical consequences of exploits.

\rqanswer{2}{
APG approaches mostly generate exploits (76\%) rather than security test cases (24\%).
In most cases (66\%), the returned artifact is \textit{incomplete}, consisting mainly of input data (i.e., the exploit payload or test input data), followed by exploit messages or other building blocks.
Incomplete test cases, in particular, may lack automated assertions.
A common method to verify a successful probe is to check the resulting program state (52\%) or inspect the execution output (19\%).
Inspecting shell spawning or program crashes is predominant for exploits.
In a few cases, the study reports no specific method, leaving the work to the user.
}

\subsection{RQ$_3$ -- Underlying Techniques and Tools}
\label{subsec:rq3_results}
\textcolor{black!75}{\textit{``\rqThree''}}

In this research question, we analyzed the main techniques (i.e., algorithms, heuristics, or procedures) and the supporting tools and frameworks that APG studies leveraged to design their own techniques.
Figure~\ref{fig:used-techniques} shows the categorization of techniques adopted across the \totalStudies surveyed studies, while Table~\ref{tbl:tools_frameworks} lists the specific tools and frameworks used as the foundation for implementing these techniques.

We classified the techniques into \textbf{ten categories}. A single study may adopt multiple technique types; thus, they are not mutually exclusive:
\begin{itemize}
	\item \textit{Formal Methods} (29 studies, 43\%). This includes constraint-solving techniques, which are popular for solving the path criteria generated during symbolic execution and producing some parts of the probe, as well as program synthesis and formal verification methods.
	\item \textit{Static Analysis} (28 studies, 42\%). These techniques are mainly used to instrument the target software's binary or source code as a preparatory step before generating exploits and security tests. Many mature program analysis frameworks (e.g., \textsc{Amoco}, \textsc{Angr}, \textsc{Javassist}, \textsc{Soot}) are used to collect vulnerability and runtime information.
	\item \textit{Symbolic Execution} (28 studies, 42\%). A symbolic execution engine is often used to synthesize path constraints required to reach a given program state. Some studies even introduce novel symbolic execution techniques specifically designed to support the generation of probes.
	\item \textit{Dynamic Analysis} (18 studies, 27\%). These techniques are often used together with static analysis to collect dynamic metadata about the target vulnerabilities and software during runtime, such as execution traces (\ref{S59}).
	\item \textit{Direct Generation} (12 studies, 18\%). Tests from models by applying mutations or direct generation from templates and grammars.
	\item \textit{Fuzzing} (12 studies, 18\%). This category refers to techniques that either leverage some existing specialized fuzzing tool or implement a custom fuzzing-like process to identify the vulnerability-triggering inputs before generating the proper exploit or security test.
	\item \textit{Search- and Optimization based} (10 studies, 15\%). This consists of using search algorithms for finding and evolving probes (often test cases). Popular search algorithms include genetic search, random search, and exhaustive search.
	\item \textit{Model-driven Analysis} (7 studies, 10\%). This category includes structured mathematical or logical models (e.g., Finite State Automata, Petri Nets) to abstract program behavior, allowing for systematic exploration of vulnerabilities.
	\item \textit{Stochastic and Learning-based Methods} (5 studies, 8\%). This category consists of techniques that leverage machine learning models trained on historical data to directly generate the desired exploit artifact.
	\item \textit{End-to-end Application Testing} (3 studies, 5\%). This includes performing end-to-end tests and vulnerability scanning to discover and trigger vulnerabilities. These are typically adopted when targeting full web applications.
\end{itemize}

\begin{figure}[t]
	\centering
	\includegraphics[width=\linewidth]{figs/rq2_used_techniques_squares.pdf}
	\caption{Underlying techniques used by APG approaches. More fine-grained categorized groups can be found in our supplementary materials~\cite{replicationpackage}.}
	\label{fig:used-techniques}
\end{figure}

\begin{table}[t]
	\centering
	\footnotesize
	\caption{Underlying tools and frameworks used by the \totalStudies surveyed studied.}
	\label{tbl:tools_frameworks}
	\begin{tabular}{p{0.3cm}|p{1.4cm}p{2cm}||p{0.3cm}|p{1.4cm}p{2cm}}
		& \textbf{Tool} & \textbf{Studies} & & \textbf{Tool} & \textbf{Studies} \\
		\midrule
		\multirow{12}{*}{\rotatebox{90}{Program Analysis}} & \textsc{Amoco} & \ref{S29}, \ref{S37} &  \multirow{5}{*}{\rotatebox{90}{Testing}} & \textsc{JUnit} & \ref{S63}\\
		& \textsc{Angr} & \ref{S29}, \ref{S32}, \ref{S36}, \ref{S38}, \ref{S40}, \ref{S47}, \ref{S59}  & & \textsc{Robotium} & \ref{S27}\\
		& \textsc{PIN} & \ref{S07}, \ref{S09}   & & \textsc{RestTestGen} & \ref{S60}\\
		& \textsc{Soot} & \ref{S26}  & & \textsc{Selenium} & \ref{S24}, \ref{S66}\\
		& \textsc{Javassist} & \ref{S67}  & & \textsc{SoapUI} & \ref{S15}\\
		\cmidrule(lr){4-6}
		
		& \textsc{ASM} & \ref{S53} & \multirow{3}{*}{\rotatebox{90}{Fuzzing}} & \textsc{Syzkaller} & \ref{S32}, \ref{S40}\\
		& \textsc{FRIDA} & \ref{S63}   &  & \textsc{AFL} & \ref{S38}, \ref{S44}\\
		& \textsc{Valgrind} & \ref{S48}   & & \textsc{Radamsa} & \ref{S48}\\
		\cmidrule(lr){4-6}
		
		& \textsc{IDA Pro} & \ref{S36}, \ref{S62}   & \multirow{3}{*}{\rotatebox{90}{Search Algo.}} & \textsc{Evosuite} & \ref{S53}, \ref{S67}\\
		& \textsc{ANTLR} & \ref{S12}, \ref{S47}, \ref{S61}   &  & \textsc{JMetal} & \ref{S63}\\
		\cmidrule(lr){1-3}
		\multirow{8}{*}{\rotatebox{90}{Symbolic Execution}} & \textsc{KLEE} & \ref{S06}, \ref{S40}, \ref{S44}, \ref{S48}, \ref{S57}   & & \textsc{GALib} & \ref{S03}\\
		& \textsc{S2E} & \ref{S13}, \ref{S40}, \ref{S65}   & & \textsc{DEAP} & \ref{S31}\\
		\cmidrule(lr){4-6}
		
		& \textsc{Crest} & \ref{S19} & \multirow{4}{*}{\rotatebox{90}{ML}} & \textsc{PyTorch} & \ref{S55} \\
		& \textsc{Angr} & \ref{S29}, \ref{S32}, \ref{S36}, \ref{S47}, \ref{S59} & & \textsc{OpenNMT} & \ref{S58} \\
		& \textsc{SPF} & \ref{S41} & & \textsc{XNMT} & \ref{S49} \\
		& \textsc{DalvikVM} & \ref{S30} & & \textsc{CoreNLP} & \ref{S65} \\
		\cmidrule(lr){4-6}
		
		& \textsc{Narcissus} & \ref{S22} & \multirow{6}{*}{\rotatebox{90}{Localization (Sup.)}} & \textsc{EBDS} & \ref{S02}\\
		\cmidrule(lr){1-3}
		\multirow{6}{*}{\rotatebox{90}{Solver (Sup.)}} & \textsc{Z3} & \ref{S07}, \ref{S09}, \ref{S18}, \ref{S20}, \ref{S21}--\ref{S22}, \ref{S26}, \ref{S30}, \ref{S33}, \ref{S35}, \ref{S37}, \ref{S40}, \ref{S44}, \ref{S47}, \ref{S57}, \ref{S59}, \ref{S61}  &  & \textsc{Splint} & \ref{S03}\\
		& \textsc{Z3-Str} & \ref{S20}--\ref{S22}, \ref{S30} & & \textsc{CPPChecknt} & \ref{S10}\\
		& \textsc{Hampi} & \ref{S04}, \ref{S23} & & \textsc{TAPS} & \ref{S22}\\
		\cmidrule(lr){4-6}
		
		& \textsc{STP} & \ref{S06} \\
		\cmidrule(lr){1-3}
		 
		 \multirow{3}{*}{\rotatebox{90}{Emu. (Sup.)}} & \textsc{QEMU} & \ref{S32}, \ref{S36}, \ref{S38}\\
		& \textsc{Android Emulator} & \ref{S28}\\
		& \textsc{VMWare} & \ref{S39}\\
		\cmidrule(lr){1-3}
	\end{tabular}
\end{table}

We note that studies in the \textit{Fuzzing} category reused an existing fuzzing tool to support the generation process, making at most minor modifications.
Indeed, these studies did not have the main goal of introducing a new fuzzing algorithm or method, but to use it as one element of the probe-generation pipeline.
This aligns with our defined scope (Section~\ref{subsub:exclusion}).

\subsubsection{Formal Methods}
This was the most widespread category, with 29 studies involved, and it mainly comprises constraint-solving methods, which are usually tied to techniques that also use symbolic execution (and thus belong to that separate category).
We further characterized it into three distinct subtypes:

\begin{itemize}
	\item \textit{Constraint Solving} (28 studies): The dominant subtype, mainly used for resolving execution path constraints and, therefore, synthesizing an artifact (often, a payload). This is further divided into \textit{Standard Constraint Solving} (22 studies) and \textit{String Constraint Solving} (8 studies). The former concerns solving formulas involving symbolic variables (often from formal parameters or global variables in a program), while the latter solves formulas concerning specifically string inputs.
	\item \textit{Program Synthesis} (2 studies): This technique is employed to automatically generate code snippets or execution sequences that satisfy pre-specified security specifications.
	\item \textit{Formal Verification} (1 study): This only covers the weakest precondition analysis made by Brumley et al. (\ref{S02}) for deriving the exact mathematical conditions required to trigger software vulnerabilities.
\end{itemize}

\subsubsection{Static Analysis}
\textit{Static Analysis} was the second most diffused category, used in 28 distinct studies, and collects vulnerability and software metadata without code execution.
This category encompasses plenty of specialized techniques:

\begin{itemize}
	\item \textit{Data Flow Analysis} (11 studies): The most prevalent subtype, used to track how information propagates through program variables, which includes niche techniques like \textit{Intent Flow Analysis} (\ref{S27}).
	\item \textit{Static Taint Analysis} (9 studies): This is used to isolate data-flow paths starting from untrusted or tainted user inputs to locate unsafe sinks.
	\item \textit{Control Flow Analysis} (8 studies): This technique reconstructs execution paths, control dependencies, and structural logic blocks across the target software.
	\item \textit{Call Graph Analysis} (5 studies): This is used to connect how procedures (subroutines) call one another in an application.
    \item \textit{Static Binary Analysis} (5 studies): This is used to inspect compiled binaries directly when access to source code is restricted.
	\item \textit{Program Slicing} (4 studies): This narrows down and isolates minimal program subsets that directly influence specific security-critical statements.
	\item \textit{Unified Program Graph Analysis} (2 studies): This sub-category incorporates advanced multi-dimensional graph models extracted from code, so it includes \textit{Program Dependency Analysis} (1 study), \textit{Code Property Graph Analysis} (1 study), and \textit{System Dependence Graph} (1 study).
    \item \textit{Type Inference Analysis} (1 study): Do et al. (\ref{S19}) insert runtime violation checks to perform pointer-level type analysis, aiming to detect potential buffer overflow vulnerabilities that are later targeted using the concolic execution framework \textsc{Crest}.
\end{itemize}

Overall, this reflects a typical workflow for automatically generating vulnerability exploits: (1) the vulnerable program is statically instrumented and analyzed using these graph or data-flow frameworks to locate the vulnerability; (2) a symbolic execution engine explores exploitable states, constructs path constraints, and generates exploit inputs by solving those constraints via a solver.
From Table~\ref{tbl:tools_frameworks}, we observe that \textsc{Angr} and \textsc{Antlr} are the most widely used tools for performing static and taint analysis of binaries.
In particular, \textsc{Antlr} is also frequently used to build and parse custom grammar-based specifications, e.g., the Exploit Specification Language (ESL) in the work by Muntean et al. (\ref{S47}), while \textsc{Angr} can also perform symbolic execution.

\subsubsection{Symbolic Execution}
On par with static analysis, symbolic execution is also a widespread category, adopted by 28 studies.
\begin{itemize}
	\item \textit{Standard symbolic execution} (19 studies): This follows the traditional approach, which does not run the program but instead executes it symbolically. Garcia et al. (\ref{S26}) leverage this technique to generate malicious intents (i.e., a form of payload) in Android.
	\item \textit{Concolic execution} (9 studies): Also known as ``Dynamic symbolic execution'' (DSE), this combines standard symbolic execution and program executions with concrete inputs to give guidance and confirmations to the path exploration made by the former. For instance, Kieyzun et al. (\ref{S04}) utilize this technique in their tool, \textsc{Ardilla}, to automatically create SQL injection and cross-site scripting (XSS) attacks.
	\item \textit{Selective symbolic execution} (2 studies): This performs the symbolic execution only on specific components of interest while executing the rest of the system concretely to reduce complexity. For example, in the study of Android hybrid apps (\ref{S30}), this technique is used to bypass complex WebView internals and focus on the JavaScript-to-native bridge to automatically generate event-oriented exploits.
\end{itemize}

From Table~\ref{tbl:tools_frameworks}, we observe that \textsc{Klee}~\cite{cadar2008klee} and \textsc{Angr} emerge as the predominant symbolic execution engines, each used in 5 studies.
Symbolic execution engines often work in conjunction with constraint solvers, with \textsc{Z3}, the most popular, used in 17 studies.
For more complex input data types, such as strings that \textsc{Z3} could not handle, \textsc{Z3-Str}~\cite{zheng2013z3} (4 studies) and \textsc{Hampi}~\cite{kiezun2009hampi} (2 studies) are used instead.

\subsubsection{Dynamic Analysis}
This second core category spans 18 studies and involves executing the target program to analyze its behavior, runtime state, and execution properties.

\begin{itemize}
	\item \textit{Execution Trace Analysis} (10 studies): The primary sub-category involves inspecting program execution paths during runtime. This technique includes specialized variants such as \textit{Sample Execution Analysis} (1 study) and \textit{Dynamic Control Flow Analysis} (1 study).
	\item \textit{Dynamic Taint Analysis} (5 studies): This technique is frequently employed to track data flow, monitoring how untrusted data propagates through variables to reach security-critical sinks. Unlike its static equivalent (part of the ``Static Analysis'' category), this includes concrete executions.
	\item \textit{Runtime Monitoring} (3 studies): Focuses on observing system-level operations or execution environments during runtime to detect anomalies.
\end{itemize}

\subsubsection{Direct Generation}
\label{subsec:model-based-generation}

This category is characterized by methods that can directly generate the intended artifact, comprising three sub-categories:
\begin{itemize}
	\item \textit{Mutation-based Generation} (7 studies): The most prominent subtype, which takes existing seed inputs or baseline payloads, then iteratively applies mutation operators and sends the mutated candidates to the target to trigger vulnerable program states.
	\item \textit{Template-based Generation} (4 studies): This technique relies on predefined templates to describe how the desired inputs, payloads, or test cases should structurally be. In particular, the study by Chaleshtari et al. (\ref{S63}) converts pre-made metamorphic relations (written in a domain-specific language) into executable Java programs that test for specific vulnerabilities in web applications.
	\item \textit{Grammar- and Rule-based Generation} (3 studies): These techniques focus on generating inputs that strictly conform to protocol or format restrictions, split into \textit{Grammar-based Generation} (2 studies) and \textit{Rule-based Generation} (1 study).
\end{itemize}

\subsubsection{Fuzzing}
12 studies use pre-existing fuzzing tools or implement a process that follows the general idea to generate useful vulnerability-triggering inputs, which are later used to construct the desired exploit or test.
We identified two subtypes based on the fuzzing approach implemented: \textit{black-box} (4 studies, 6\%) and \textit{grey-box} (8 studies, 12\%).
The former refers to approaches that generate random, malformed, and mutated inputs without any knowledge of the target's internal structure, relying only on external signals (mainly crashes, hangs, or error messages).
The latter instead uses lightweight \textit{code instrumentation} to track which branches and code paths each input exercises, guiding mutations to reach new code areas. 

In most cases, approaches leveraging fuzzing are constrained to the type of software they can target.
For example, AFL typically targets command-line programs (i.e., binaries), while    \textsc{Syzkaller} supports OS kernels.
Yet, this does not mean that fuzzing cannot be applied to other domains.
For example, \ref{S52} implements a grey-box fuzzing method to generate exploits for PHP object injection (POI) vulnerabilities; this requires the PHP application's source code.
\ref{S45}, instead, implements fuzzing for APIs (OpenAPI, Swagger) and uses a custom black-box approach.
In this regard, there are fewer available tools that implement pure black-box fuzzing.
One of these is \textsc{Radamsa}, which adopts more of a ``random testing'' approach, as its input sampling relies heavily on randomness and it uses no feedback from the program.

Lastly, we observed that some APG approaches gave a custom name to the fuzzing method they implemented, e.g., Layout-oriented Fuzzing (\ref{S38}) and Capability-guided Fuzzing (\ref{S40}).

\subsubsection{Search- and Optimization-based} These techniques usually employ a search algorithm to find candidate probes according to some objective functions (modeled after an exploitation goal or a vulnerability trigger event).
\begin{itemize}
	\item \textit{Evolutionary Algorithms} (7 studies): The most prevalent subtype, which models potential inputs as chromosomes and applies selection, crossover, and mutation operators to evolve a population of candidate probes. For example, Kang et al. (\ref{S53}) proposed a technique, called \textsc{Transfer}, to evolve a population of JUnit test cases to trigger a vulnerable state caused by a vulnerability in a third-party library.
    The harness to run an evolution algorithm loop and run the test cases is provided by tools like \textsc{Evosuite}~\cite{Fraser:evosuite:2013}.
	\item \textit{Heuristic Search} (2 studies): Some studies design their own heuristic search strategies to fit their exploit generation problems. For example, Heelan (\ref{S31}) introduces \textit{pseudo-random black box search} to find inputs that align the buffer vulnerabilities with target heap objects, automating the heap layout manipulation process for exploitation.
    \item \textit{Linear Programming} (2 studies): Studies in this subcategory often model the problem using linear programming and then solve it to find the exploits. For example, Li et al. (\ref{S64}) try to model the heap layout constraints such as filling target holes and satisfying program execution logic as an integer linear programming problem, which is later solved using Z3~\cite{zheng2013z3}.
    \item \textit{Exhaustive Search} (1 study): Yilmaz et al. (\ref{S39}) introduce their semi-automated AEG approach \textsc{GuidExp}, which relies on exhaustive search (rather than fuzzers or symbolic execution) to explore execution paths and synthesize exploit scripts for ActionScript Virtual Machine (AVM) vulnerabilities.
\end{itemize}

\subsubsection{Model-driven Analysis}

\textit{Model-driven Analysis} (7 studies) primarily comprises state-transition and behavioral modeling methods that are used to evaluate system executions in the abstract. In practice, this means constructing formal representations of software behavior.

Within this category, we identified two types: \textit{Finite State Automata Analysis} (6 studies) and \textit{Petri Net Analysis} (1 study).
The former consists of modeling system behaviors and input constraints through finite state machines to verify valid state transitions, ensure protocol compliance, and detect unreachable or erroneous states within a defined lifecycle.
Petri Net Analysis instead offers specialized capabilities for modeling concurrent or parallel system behaviors to analyze distributed workflows and synchronization; this technique utilizes a graphical and mathematical framework involving places, transitions, and tokens to rigorously examine asynchronous processes, identify potential deadlocks, and manage resource contention in complex distributed environments.

\subsubsection{Stochastic and Learning-based Methods} 
Recently, researchers have begun using stochastic and learning-based methods to learn and directly generate exploits or security tests.
We subcategorized this group into \textit{Seq2Seq-based} approaches (4 studies) and \textit{Reinforcement Learning} (with one study).
The former use Seq2Seq models to transform partial inputs or textual prompts into probes, typically trained in a supervised manner.
The latter rely on reward signals to improve the generation policy.

For example, Liguori et al. (\ref{S49}) and Yang et al. (\ref{S58}) developed neural machine translation-based approaches to generate exploit payloads (i.e., not fully working exploits) in Assembly and Python given an intent expressed in natural language.
Liu et al. (\ref{S43}) trained a model to learn the semantics of SQL injection attacks from benign inputs.
In such studies, the generated probes were not evaluated or confirmed to be effective; instead, they were evaluated on a ground-truth dataset using metrics such as precision, recall, and BLEU, rather than run in practice.

On the other hand, Yao et al. (\ref{S66}) generate XSS attack vectors as a step-by-step decision process in which a reinforcement learning algorithm (an improved \textit{Dueling DDQN}) repeatedly picks mutation strategies to apply to the initial vector’s mutation points, with an edit-distance-based reward guiding it toward better mutations, and the resulting vectors are then checked for effectiveness using static semantic analysis.

\subsubsection{End-to-end Application Testing}
This last category consists of application-layer testing methods that treat the target as a black box.
The first subtype is \textit{Browser-driven Testing} (2 studies), which simulates user interactions (e.g., clicks) with end-to-end tests implemented through web drivers like \textsc{Selenium}.
The second is \textit{Vulnerability Scanning-like} (2 studies).
These approaches either analyze the website's \textit{navigation path} to understand what HTTP requests are needed to explore specific paths, or run existing dynamic application security testing (DAST) tools such as OWASP ZAP.
In both cases, they then post-process the results to produce probes.

\rqanswer{3}{
APG studies mostly rely on \textit{Formal Methods} (43\%), \textit{Static Analysis} (42\%), and \textit{Symbolic Execution} (42\%) to generate exploits.
Constraint solving (a formal method) is an essential element for solving the path constraints obtained from symbolic execution.
Static analysis, instead, serves as a preparatory step rather than the algorithm that generates the probe artifact.
Conversely, other studies use Model-based Generation (8\%), Fuzzing (8\%), and Search- and Optimization-based techniques (7\%) to generate security tests and find inputs that trigger vulnerabilities.
Additionally, some APG techniques such as Learning-based Methods (3\%) directly synthesize payloads using learning models.
}

\subsection{RQ$_4$ -- Input Requirements}

\textcolor{black!75}{\textit{``\rqFour''}}

\begin{table}[t]
\centering
\caption{Types of inputs required by the approaches introduced in the \totalStudies surveyed studies. Source code can have three forms: \entrypoint = Entry Point; \method = Method or Function; \fragment = Code Fragment. A dash (--) means that no auxiliary input is required.}
\label{tbl:rq4-input}
\resizebox{0.98\linewidth}{!}{%
\begin{tabular}{c|ccc|p{4.4cm}||c|ccc|p{4.4cm}}
\hline
\textbf{Study} & \multicolumn{3}{c|}{\textbf{Target Input}} & \textbf{Auxiliary Input} & \textbf{Study} & \multicolumn{3}{c|}{\textbf{Main Inputs}} & \textbf{Auxiliary Input} \\
\cline{2-4} \cline{7-9}
 & \rotheader{Source Code} & \rotheader{Executable} & \rotheader{Specification } & & & \rotheader{Source Code} & \rotheader{Executable} & \rotheader{Specification } & \\
\hline
\ref{S01} &   & \tick &   & PoC/Payload & \ref{S35} &   & \tick &  & Vulnerability-Triggering Input \\
\ref{S02} &   & \tick &   & Patched Target & \ref{S36} &   & \tick &   & PoC/Payload; Baseline Input \\
\ref{S03} & \entrypoint &   &   & -- & \ref{S37} &   & \tick &   & -- \\
\ref{S04} & \entrypoint &   &   & Attack Specification & \ref{S38} &   & \tick &   & PoC/Payload \\
\ref{S05} &   & \tick &   & -- & \ref{S39} &   & \tick &   & PoC/Payload; Attack Specification \\
\ref{S06} &   & \tick &   & -- & \ref{S40} &   & \tick &   & PoC/Payload \\
\ref{S07} &   & \tick &   & -- & \ref{S41} &   & \tick &   & -- \\
\ref{S08} & \entrypoint &   &   & -- & \ref{S42} & \entrypoint &   &   & Existing Tests \\
\ref{S09} &   & \tick &  & Attack Specification & \ref{S43} &   & \tick &   & Baseline Input \\
\ref{S10} & \entrypoint &   &   & Attack Specification, Execution Traces & \ref{S44} &   & \tick &   & -- \\
\ref{S11} &   & \tick &   & Execution Traces & \ref{S45} &   & \tick & \tick & -- \\
\ref{S12} & \method &   &   & -- & \ref{S46} &   & \tick &   & Baseline Input; Attack Specification \\
\ref{S13} &   & \tick &  & Vulnerability-Triggering Input & \ref{S47} &   & \tick &  \tick & Attack Specification \\
\ref{S14} &   & \tick &   & Attack Specification & \ref{S48} & \entrypoint & \tick &   & SAST Warnings \\
\ref{S15} &   & \tick & \tick & Baseline Input & \ref{S49} &   &   &   & Natural language intent \\
\ref{S16} & \entrypoint &   &   & -- & \ref{S50} & \fragment &   &   & -- \\
\ref{S17} &   & \tick &   & -- & \ref{S51} & \fragment &   &   & -- \\
\ref{S18} &   & \tick & & Vulnerability-Triggering Input, Baseline Input & \ref{S52} & \entrypoint & \tick &   & -- \\
\ref{S19} & \entrypoint &   &   & -- & \ref{S53} & \entrypoint &   &   & Vulnerability-Witnessing Test \\
\ref{S20} & \entrypoint &   &   & -- & \ref{S54} & \entrypoint &   & \tick & Vulnerability-Triggering Input \\
\ref{S21} & \entrypoint &   &   & -- & \ref{S55} & \fragment &   &   & -- \\
\ref{S22} & \entrypoint & \tick &   & Attack Specification & \ref{S56} & \entrypoint &   &   & Existing Tests \\
\ref{S23} &   & \tick & \tick & Baseline Input & \ref{S57} &   & \tick &   & -- \\
\ref{S24} &   & \tick &   & -- & \ref{S58} &   &   &   & Natural-Language Intent \\
\ref{S25} &   & \tick &   & -- & \ref{S59} &   & \tick &   & PoC/Payload \\
\ref{S26} &   & \tick &   & -- & \ref{S60} &   & \tick & \tick  & -- \\
\ref{S27} &   & \tick &   & -- & \ref{S61} & \fragment &   &   & -- \\
\ref{S28} &   & \tick &   & -- & \ref{S62} &   & \tick & \tick & -- \\
\ref{S29} &   & \tick & & Vulnerability-Witnessing Test & \ref{S63} & \entrypoint &   &   & Attack Specification \\
\ref{S30} &   & \tick &   & -- & \ref{S64} & \entrypoint &   & \tick & Vulnerability-Triggering Input \\
\ref{S31} &   & \tick &  & Vulnerability-Triggering Input, PoC/Payload, Existing Tests & \ref{S65} &   & \tick &   & PoC/Payload \\
\ref{S32} &   & \tick &   & PoC/Payload & \ref{S66} &   & \tick &   & Attack Specification \\
\ref{S33} &   & \tick &   & -- & \ref{S67} & \entrypoint &   &   & Vulnerability-Triggering Input \\
\ref{S34} & \fragment &   &   & -- & & & & & \\
\hline
\multicolumn{10}{l}{\textbf{Target Total}: Executable: \textbf{44}, Source Code: \textbf{24}, Specification: \textbf{8}.} \\
\hline
\multicolumn{10}{p{17cm}}{\textbf{Auxiliary Total (36)}: Attack Specification: \textbf{10}, PoC/Payload: \textbf{9}, Vulnerability-Triggering Input: \textbf{7}, Baseline Input: \textbf{6}, Existing Tests: \textbf{3}, Natural-Language Intent: \textbf{2}, Vulnerability-Witnessing Test: \textbf{2}, Execution Traces: \textbf{2}, Patched Target: \textbf{1}, SAST Warnings: \textbf{1}.} \\
\hline
\end{tabular}
}
\end{table}

In this research question, we analyzed the input elements APG approaches require to generate probes, summarized in Table~\ref{tbl:rq4-input}.
All approaches require at least one form of \textit{representation} of the target from which the behavior can be inferred.
We refer to this as \textit{target input}, as it concerns the target software itself.

In the majority of cases, the required form is the \textit{Executable} (44 studies, 66\%), which refers to any machine-readable binary that an operating system or a runtime can execute directly without further translation.
This suggests that many APG approaches can operate without access to the target's source code, which may make them easier to use on closed-source or third-party software.
There is, however, no guarantee that these work out of the box in all circumstances, making them difficult to use directly in CI/CD pipelines.

This is followed by \textit{Source Code} (24 studies, 36\%), which is often an \textit{entry point} (\entrypoint) of the target (18 studies), i.e., a function/method that can be called from the external (e.g., terminal, browser, or framework). This often requires providing the APG approach access to the rest of the source code reachable from that entry point.
Alternatively, the input can be a \textit{code fragment} (\fragment), i.e., a partial code snippet, a regular expression, or even an SQL statement (5 studies), or even a \textit{method or function} (\method) that implements a specific functionality, such as an input validation routine (this was found in just one study).
All APG approaches require either an executable or a source code input.
In three cases (\ref{S22}, \ref{S48}, and \ref{S52}), an APG approach requires both the executable version of the target as well as access to (part of) its source code.

A third input type is a \textit{Specification} of the target, which describes the target APIs (e.g., an OpenAPI or WSDL specification) or its memory layout.
Yet, this input is a plus, as we found no APG approach that needs only a specification.
This also applies to Model-driven approaches (see Section~\ref{subsec:rq3_results}), which we observed build the model directly from a lower-level artifact rather than asking the user for it.

In addition to the target input, many APG approaches rely on additional \textit{auxiliary inputs}, which are used to assist the probe-generation algorithm, e.g., by localizing the vulnerability or reproducing a certain vulnerable state.
Overall, 36 studies (58\%) require at least one auxiliary input, which we could classify as follows:
\begin{itemize}
	\item \textit{Attack Specification} (10 studies): a description, ruleset, profile, or constraints on the malicious input the probe is expected to deliver to the target. A typical example is an attack grammar describing the malicious input syntax.
	\item \textit{PoC/Payload} (9 studies): a pre-existing Proof-of-Concept (PoC) exploit or payload that triggers a specific vulnerable state in the target. This serves as a starting point for building a more advanced probe.
    \item \textit{Vulnerability-Triggering Input} (7 studies): an input that triggers a specific vulnerable state in the target that is integrated into the generated probe. This is similar in spirit to PoC/Payload.
	\item \textit{Baseline Input} (6 studies): a benign input passed to the target to observe its ordinary (i.e., safe) behavior. This is used to explore and understand the program behavior, which is often mutated or altered to create malicious input.
	\item \textit{Existing Tests} (3 studies): a suite of existing test cases in the target project, including those unrelated to security. These are often used to reach certain program paths or for regression.
	\item \textit{Natural Language Intent} (2 studies): a plain-text description outlining what the prospect probe is expected to behave and achieve. This is typical for Seq2Seq-based generators (see Section~\ref{subsec:rq3_results}).
	\item \textit{Vulnerability-Witnessing Test} (2 studies): a test case that can be run through a testing framework (e.g., JUnit) whose execution triggers a specific vulnerable state in the target.
	\item \textit{Execution Traces} (2 studies): step-by-step records of system behavior and logs captured during an execution run.
    \item \textit{Patched Target} (1 study): the patched version of the target. This was mainly needed to localize the vulnerability and to confirm the probe correctness (i.e., that it succeeds in the vulnerable version and fails in the patched version).
    \item \textit{SAST Warnings} (1 study): the warnings from a SAST tool. These were used in \ref{S48} to first build a reduced version of the target that has the same warnings, then use \textsc{Radamsa} and KLEE to generate the test input, and then produce the tests (in the form of \texttt{main} programs) with assertions that encode the failure location and type from the warnings.
\end{itemize}

We note that this list is not a complete taxonomy of auxiliary inputs, as a few appeared in only one or two studies, which does not allow proper generalization.

We also observed that certain auxiliary inputs can be difficult to obtain. For example, execution traces require prior offline monitoring of the target. In 10 cases (15\%), the user must provide an attack specification (e.g., a grammar); although some pre-defined specifications provided by the paper authors may be reused as is, these often must be defined manually.

In roughly one out of four cases (17 instances), the APG approach requires a ``hint'' to bootstrap the inner generation procedure, such as a PoC/Payload, a vulnerability-triggering input, or a vulnerability-witnessing test.
Effectively, these serve as an initial probe.
Consequently, the user must prepare these artifacts themselves, either manually or with support from another APG approach.
This requirement makes some APG approaches harder to deploy immediately.

Of particular interest are approaches that require no prior knowledge of the vulnerability, i.e., no artifact that reproduces a vulnerable state or describes the format of the malicious input.
This enables the APG approach to autonomously search for the vulnerable sink and generate input that exercises it. However, this autonomy may come at the cost of returning less successful probes.

\rqanswer{4}{
Most APG approaches (66\%) require the executable of the target, while only 36\% require its source code (with 3 studies requiring both).
12\% also require a specification of the target API or its memory layout, though this piece is always required alongside executable and/or source code.
75\% of the studies that require access to source code need to know the specific entry point to start the probe. 
More than half of the studies (58\%) also require auxiliary input that assists the probe generation procedure, which in 17 cases consists of an artifact that reproduces a specific vulnerable state (e.g., a PoC/Payload or a vulnerability-triggering input).
In most other cases, however, no prior knowledge of the target vulnerability is required.
}

\subsection{RQ$_5$ -- Evaluation Method}
\label{subsec:rq5}

\textcolor{black!75}{\textit{``\rqFive''}}

In this research question, we analyzed the evaluation methodologies adopted by the primary studies to assess the effectiveness of the proposed APG approaches.
In particular, we analyzed the content type, realism, programming language, and scale of the evaluation datasets.
Table~\ref{tbl:rq5-evaluation} summarizes the main characteristics of the evaluation datasets used by each of the \totalStudies surveyed studies.

\newcommand{\Yes}{\ding{51}}
\newcommand{\No}{\ding{0}}
\begingroup
\catcode`\/=\active
\gdef\activateSlashBreak{%
  \catcode`\/=\active
  \def/{\char47\allowbreak{}}%
}
 \endgroup

\begin{table*}[t]
\scriptsize
\centering
\setlength{\tabcolsep}{2pt}
\renewcommand{\arraystretch}{1.0}
\caption{Evaluation datasets and research-evaluation procedures reported by the \totalStudies surveyed studies. \textit{N/A} (not applicable) is used when the target is a regular expression or an API. The research-evaluation (RE) categories are not mutually exclusive. `Manual' indicates a manual inspection of all (`full') or selected (`sample') cases, while GT denotes a comparison against a ground truth or benchmark, which is done automatically.}
\label{tbl:rq5-evaluation}

\begingroup
\activateSlashBreak

\resizebox{0.95\linewidth}{!}{%
\begin{tabular}{c|ccp{1.2cm}|p{1.75cm}|l|ccc||c|ccp{1.2cm}|p{1.75cm}|l|ccc}
\hline

\textbf{Study} &
\multicolumn{5}{c|}{\textbf{Evaluation Dataset}} &
\multicolumn{3}{c||}{\textbf{RE}} &
\textbf{Study} &
\multicolumn{5}{c|}{\textbf{Evaluation Dataset}} &
\multicolumn{3}{c}{\textbf{RE}}
\\

\cline{2-6}
\cline{7-9}
\cline{11-15}
\cline{16-18}

&
\rotheader{Real-world} &
\rotheader{Artificial} &
\textbf{Language} &
\textbf{Target type} &
\textbf{Size} &
\rotheader{Manual (Full)} &
\rotheader{Manual (Sample) } &
\rotheader{Ground Truth} &
&
\rotheader{Real-world} &
\rotheader{Artificial} &
\textbf{Language} &
\textbf{Target type} &
\textbf{Size} &
\rotheader{Manual (Full)} &
\rotheader{Manual (Sample) } &
\rotheader{Ground Truth}
\\

\hline

\ref{S01} & \No & \Yes & C & Program & 1
& \No & \No & \No
&
\ref{S35} & \No & \Yes & C/C++ & DOP script & 4
& \No & \No & \No \\

\ref{S02} & \Yes & \No & \textit{Unsp.} & Windows program & 5
& \No & \Yes & \Yes
&
\ref{S36} & \Yes & \Yes & C & CTF/CVE prog. & 19
& \No & \Yes & \Yes \\

\ref{S03} & \Yes & \No & C/C++ & Function & 7
& \No & \No & \No
&
\ref{S37} & \Yes & \Yes & C & CTF/CVE prog. & 9
& \No & \No & \No \\

\ref{S04} & \Yes & \No & PHP & Web app & 5
& \Yes & \No & \No
&
\ref{S38} & \Yes & \Yes & C & CTF/CVE prog. & 24
& \No & \No & \No \\

\ref{S05} & \Yes & \No & \textit{Unsp.} & Linux OS & 4
& \No & \No & \No
&
\ref{S39} & \Yes & \No & ActionS. & AVM vul & 11
& \No & \No & \No \\

\ref{S06} & \Yes & \No & C/C++ & Program & 14
& \No & \No & \No
&
\ref{S40} & \Yes & \No & C & Linux kernel vul & 10
& \No & \Yes & \No \\

\ref{S07} & \Yes & \No & C & Network app & 3
& \No & \No & \No
&
\ref{S41} & \Yes & \No & Java & Android version & 7
& \No & \No & \No \\

\ref{S08} & \Yes & \No & PHP & Web app & 6
& \Yes & \No & \No
&
\ref{S42} & \No & \Yes & Java & Synthetic benchmark/Real app & 4/3
& \No & \No & \No \\

\ref{S09} & \Yes & \No & \textit{Unsp.} & Linux/Win. prog. & 22/7
& \No & \No & \No
&
\ref{S43} & \Yes & \No & Java & Web app & 6
& \No & \No & \No \\

\ref{S10} & \Yes & \Yes & Java, PHP, C/C++ & Web app/Program & 3/2/1
& \No & \Yes & \Yes
&
\ref{S44} & \Yes & \Yes & C/C++ & Program & 12
& \Yes & \No & \Yes \\

\ref{S11} & \Yes & \No & \textit{Unsp.} & Custom OS & 1
& \No & \No & \No
&
\ref{S45} & \Yes & \No & N/A & Azure/Office365 & 2/1
& \No & \No & \No \\

\ref{S12} & \Yes & \Yes & PHP & Function & 15,552
& \No & \No & \No
&
\ref{S46} & \Yes & \No & PHP & Web app & 33
& \Yes & \No & \No \\

\ref{S13} & \Yes & \Yes & C & Program & 38
& \No & \No & \No
&
\ref{S47} & \Yes & \No & \textit{Unsp.} & Program & 7
& \No & \No & \No \\

\ref{S14} & \Yes & \Yes & PHP & DVWA/Real app & 1/5
& \No & \No & \No
&
\ref{S48} & \Yes & \No & C/C++ & Program & 12
& \No & \Yes & \Yes \\

\ref{S15} & \Yes & \No & PHP & SOAP WS & 2
& \Yes & \No & \No
&
\ref{S49} & \Yes & \No & Python ASM & Program & 20/20
& \Yes & \No & \No \\

\ref{S16} & \Yes & \No & PHP & Web app & 10
& \No & \No & \Yes
&
\ref{S50} & \Yes & \No & N/A & Regex & 29,088
& \No & \No & \No \\

\ref{S17} & \Yes & \No & \textit{Unsp.} & Web app & 17,913
& \No & \Yes & \No
&
\ref{S51} & \Yes & \No & N/A & Regex & 609,992
& \No & \No & \No \\

\ref{S18} & \Yes & \No & C/C++ & Ubuntu/Win. prog. & 7/1
& \No & \No & \Yes
&
\ref{S52} & \Yes & \Yes & PHP & Custom/Real app & 12/17
& \Yes & \No & \Yes \\

\ref{S19} & \Yes & \No & C/C++ & Program & 7
& \No & \No & \Yes
&
\ref{S53} & \Yes & \No & Java & Client app & 64
& \Yes & \No & \Yes \\

\ref{S20} & \Yes & \No & PHP & Exploit kit & 16
& \No & \No & \No
&
\ref{S54} & \Yes & \No & C/C++ & Linux kernel vuls & 22
& \No & \No & \No \\

\ref{S21} & \Yes & \No & PHP & Web app & 6
& \Yes & \No & \No
&
\ref{S55} & \Yes & \Yes & Assembly & Shellcode & 320
& \No & \Yes & \No \\

\ref{S22} & \Yes & \No & PHP & Web app & 26
& \Yes & \No & \No
&
\ref{S56} & \Yes & \No & PHP & Template engine & 7
& \Yes & \No & \No \\

\ref{S23} & \Yes & \Yes & PHP & Web app & 9
& \No & \Yes & \No
&
\ref{S57} & \Yes & \No & C++/WASM & Smart contract & 162
& \No & \No & \Yes \\

\ref{S24} & \Yes & \No & \textit{Unsp.} & SOAP WS & 44
& \No & \Yes & \No
&
\ref{S58} & \Yes & \No & Python ASM & Program & 20/20
& \Yes & \No & \No \\

\ref{S25} & \Yes & \Yes & \textit{Unsp.} & XMLMao/Real app & 1/4
& \No & \No & \No
&
\ref{S59} & \Yes & \Yes & C & CTF/CVE prog. & 16
& \No & \Yes & \No \\

\ref{S26} & \Yes & \No & \textit{Unsp.} & MPWAs & 507
& \Yes & \No & \Yes
&
\ref{S60} & \Yes & \Yes &
Java/Python/ASP.NET/Node.js &
REST APIs & 15
& \No & \No & \Yes \\

\ref{S27} & \Yes & \No & Java & Android version & 11
& \No & \Yes & \No
&
\ref{S61} & \Yes & \No & N/A & Regex & 353,012
& \Yes & \No & \Yes \\

\ref{S28} & \Yes & \No & Java & Android app & 5/10k
& \No & \Yes & \No
&
\ref{S62} & \Yes & \No & JavaS., WXML & API discrepancy & 214
& \No & \No & \No \\

\ref{S29} & \Yes & \No & Java & Android app & 20k
& \No & \No & \Yes
&
\ref{S63} & \Yes & \No & Java/PHP & Web app & 1/1
& \Yes & \No & \Yes \\

\ref{S30} & \Yes & \No & Java, JavaS. & Android app & 13k
& \Yes & \No & \No
&
\ref{S64} & \Yes & \No & C/C++ Python PHP & Program & 12
& \Yes & \No & \Yes \\

\ref{S31} & \Yes & \No & PHP & Interpreter bug & 50
& \No & \No & \No
&
\ref{S65} & \Yes & \Yes & \textit{Unsp.} & CTF/Real prog. & 96/3
& \No & \No & \No \\

\ref{S32} & \Yes & \No & C/C++ & Linux kernel UAF vul & 15
& \No & \No & \Yes
&
\ref{S66} & \Yes & \Yes & \textit{Unsp.} & DVWA/Real app & 1/3
& \No & \Yes & \No \\

\ref{S33} & \Yes & \No & Solidity EVM & ETH smart contract & 38,757
& \No & \No & \No
&
\ref{S67} & \Yes & \No & Java & Client app & 60
& \No & \No & \Yes \\

\ref{S34} & \Yes & \No & N/A & Regex & 29,088
& \No & \No & \No
&
& & & & & & & & \\

\hline

\textbf{Tot.} &
\multicolumn{17}{l}{
Real-world: \textbf{64}, Artificial: \textbf{19}
}
\\

\hline

\textbf{RE tot.} &
\multicolumn{17}{l}{
Manual (full): \textbf{17}, Manual (sample): \textbf{14}, Ground Truth: \textbf{20}
}
\\

\hline

\multicolumn{18}{l}{
\textit{RE = Research evaluation. \textit{Unsp.} = Unspecified.}
}
\\
\hline
\end{tabular}
}
\endgroup
\end{table*}

Real-world targets are used by 64 of the 67 studies (96\%): 48 studies rely exclusively on them, while 16 combine them with artificial targets.
Only three studies (\ref{S01}, \ref{S35}, and \ref{S42}) evaluate their approaches exclusively on artificial targets.

Artificial datasets are typically used to assess whether an approach can generate exploits or security tests for known or injected vulnerabilities under controlled conditions.
Real-world datasets are instead used to assess the practical applicability and robustness of the approach in realistic scenarios.
The combination of \textit{in-vitro} and \textit{in-vivo} evaluation targets provides stronger empirical evidence while partially mitigating the limited availability of large-scale real-world vulnerability datasets.
For example, studies such as \ref{S14}, \ref{S25}, and \ref{S66} combine intentionally vulnerable benchmark applications, such as DVWA or XMLMao, with real-world PHP applications.

The evaluation targets used across the studies are highly heterogeneous and span multiple domains, including Linux kernels, operating systems, binary and source-code programs, web applications, Android applications, web services, smart contracts, APIs, shellcodes, and regular expressions.
After mapping the descriptions reported in the primary studies to the target-type categories used in Table~\ref{tbl:rq5-evaluation}, programs and web applications are the most frequent target types, appearing in 10 studies (14.9\%) each.
Regular expressions and CTF/CVE programs are each used in 4 studies (6.0\%), whereas Android applications are used in 3 studies (4.5\%).
Most studies focus on application- or program-level evaluation targets, whereas only a limited number evaluate their approaches on fine-grained artifacts such as functions, DOP scripts, or regular expressions.

Regarding the programming languages of the evaluation targets, C and C++ are the most frequently represented languages, appearing in 20 studies (29.9\%), followed by PHP in 17 studies (25.4\%) and Java in 12 studies (17.9\%). Other languages occur only occasionally. These counts are not mutually exclusive, since some studies evaluate targets implemented in multiple programming languages. For 11 studies (16.4\%), the programming language is unspecified because it is not reported clearly in the paper, while for 5 studies (7.5\%) it is not applicable because the evaluated target is a language-independent artifact, such as a regular expression or an API-level target.

The size of the evaluation datasets varies substantially depending on the complexity and granularity of the evaluation targets.
Studies targeting Linux kernels, operating systems, or low-level vulnerabilities typically evaluate their approaches on relatively small datasets, often involving fewer than 10 subjects, due to the complexity and engineering effort required to analyze such systems.
Web-application and program-level studies generally use datasets ranging from a few subjects to a few dozen subjects.
Finally, evaluations involving lightweight artifacts such as regular expressions, shellcodes, or code fragments often scale to tens or hundreds of thousands of targets, benefiting from the availability of large public corpora and automated generation procedures.

In RQ$_2$, we mapped the verification oracle used by the APG approaches.
Here, we analyzed whether the authors performed an additional \textbf{validation} to assess the quality of the generated probes.
These procedures complement the verification oracle by providing additional evidence for the credibility of the reported empirical results.
Overall, 39 studies (58.2\%) report the results of at least one additional procedure.
These procedures include manual inspections of either the entire set or a subset of generated artifacts, results, or evaluation cases.
In particular, 17 studies (25.4\%) manually inspected the entire set, whereas 14 (20.9\%) inspected only a sample.
The latter category may consist of a fixed-size random selection of cases or a selection based on a specific criterion, e.g., failures or other interesting cases.
In contrast, 20 studies (29.9\%) compared their results against benchmarks, ground truth, or external references, such as known CVEs, public proof-of-concept exploits, or previously reported attacks.
The categories are not mutually exclusive, as some studies performed more than one evaluation.

\rqanswer{5}{
APG evaluations rely mostly on real-world datasets and are often complemented by artificial benchmarks to test approaches in controlled settings.
Target domains are highly diverse, ranging from low-level systems (Linux kernels, OS) and web/mobile applications to APIs and smart contracts.
Dataset size depends on subject complexity: complex targets (e.g., kernels) use small datasets (fewer than 10 subjects), whereas lightweight targets (e.g., regular expressions, code fragments) reach hundreds of thousands.
Beyond automated execution oracles, 58\% of studies (39 out of 67) perform additional research-evaluation procedures, using full manual checks (17 studies), sample/selected-case checks (14 studies), or comparisons against external ground truth and CVEs (20 studies).
}

\subsection{RQ$_6$ -- Implementation and Tool Availability}
\label{subsec:rq6}

\textcolor{black!75}{\textit{``\rqSix''}}

In this research question, we analyzed the tools implementing APG approaches in the surveyed papers, focusing on their level of automation, suitability for integration into CI/CD pipelines, type of released artifact, runtime platform, documentation, availability, and deployment support.
Table~\ref{tbl:avail-tools} summarizes this information for the accessible tools.

\begin{table}[t]
\centering
\scriptsize
\caption{Publicly available tools identified in the reviewed studies (verified as of September 23, 2026). The automation level (\textit{Autom.}) can be \textit{Semi-automated} (\ding{71}) or \textit{Fully-automated} (\ding{70}). \textit{CI/CD} indicates the apparent suitability for integration into CI/CD pipelines with limited additional engineering effort. \textit{Type} indicates whether the release is a Standalone Tool (\textit{ST}), Research Prototype (\textit{RP}), or Extensible Framework (\textit{EF}).}
\label{tbl:avail-tools}
\resizebox{\textwidth}{!}{%
\begin{tabular}{l|c|c|c|c|c|c|c}
    \hline
    \textbf{Tool (Study)} & \textbf{Autom.} & \textbf{CI/CD} & \textbf{Type} &
    \textbf{Runtime} & \textbf{Docs} & \textbf{Deploy.} & \textbf{Link} \\
    \hline

    \textsc{Mista} (\ref{S10})
    & \ding{71} & \ding{55} & EF & Java & \ding{55} & --
    & \href{https://github.com/dianxiangxu/MISTA}{GitHub} \\

    \textsc{SSOScan} (\ref{S17})
    & \ding{70} & \ding{55} & ST & Ruby & \ding{51} & --
    & \href{https://web.archive.org/web/20260416020635/http://ssoscan.org/}{Wayback} \\

    \textsc{Navex} (\ref{S22})
    & \ding{70} & \ding{55} & ST & Java & \ding{51} & --
    & \href{https://github.com/aalhuz/navex}{GitHub} \\

    \textsc{Shrike} (\ref{S31})
    & \ding{71} & \ding{55} & RP & Python & \ding{51} & --
    & \href{https://github.com/SeanHeelan/HeapLayout}{GitHub} \\

    \textsc{Fuze} (\ref{S32})
    & \ding{70} & \ding{55} & RP & Python & \ding{51} & --
    & \href{https://github.com/ww9210/Linux_kernel_exploits}{GitHub} \\

    \textsc{Teether} (\ref{S33})
    & \ding{70} & \ding{51} & ST & Python & \ding{51} & --
    & \href{https://github.com/nescio007/teether}{GitHub} \\

    \textsc{ReScue} (\ref{S34})
    & \ding{70} & \ding{51} & ST & Java & \ding{51} & --
    & \href{https://2bdenny.github.io/ReScue/}{Web} \\

    \textsc{Kepler} (\ref{S36})
    & \ding{71} & \ding{55} & RP & Python & \ding{51} & --
    & \href{https://github.com/ww9210/kepler-cfhp}{GitHub} \\

    \textsc{ARG} (\ref{S37})
    & \ding{70} & \ding{55} & RP & Python & \ding{55} & --
    & \href{https://github.com/wy666444/auto_rop}{GitHub} \\

    \textsc{Koobe} (\ref{S40})
    & \ding{70} & \ding{55} & RP & Python & \ding{51} & --
    & \href{https://github.com/seclab-ucr/KOOBE}{GitHub} \\

    \textsc{Centaur} (\ref{S41})
    & \ding{70} & \ding{55} & EF & Java & \ding{51} & --
    & \href{https://github.com/Android-Framework-Symbolic-Executor/Centaur}{GitHub} \\

    \textsc{Rivulet} (\ref{S42})
    & \ding{70} & \ding{51} & ST & Java & \ding{51} & VM
    & \href{https://github.com/gmu-swe/rivulet}{GitHub} \\

    \textsc{DeepSQLi} (\ref{S43})
    & \ding{70} & \ding{51} & ST & Java & \ding{51} & --
    & \href{https://github.com/COLA-Laboratory/issta2020}{GitHub} \\

    \textsc{Savior} (\ref{S44})
    & \ding{70} & \ding{51} & EF & LLVM & \ding{51} & Docker
    & \href{https://github.com/evanmak/savior-source}{GitHub} \\

    \textsc{Fuse} (\ref{S46})
    & \ding{71} & \ding{51} & ST & Python & \ding{51} & --
    & \href{https://github.com/WSP-LAB/FUSE}{GitHub} \\

    \textsc{iTOP} (\ref{S47})
    & \ding{70} & \ding{51} & EF & Python & \ding{51} & --
    & \href{https://github.com/TeamVault/iTOP-clone}{GitHub} \\

    \textsc{EVIL} (\ref{S49})
    & \ding{71} & \ding{55} & ST & Python & \ding{51} & --
    & \href{https://github.com/dessertlab/EVIL}{GitHub} \\

    \textsc{Revealer} (\ref{S50})
    & \ding{70} & \ding{51} & ST & Java & \ding{51} & --
    & \href{https://github.com/cuhk-seclab/Revealer}{GitHub} \\

    \textsc{Fugio} (\ref{S52})
    & \ding{71} & \ding{55} & RP & Python & \ding{51} & Docker
    & \href{https://github.com/WSP-LAB/FUGIO}{GitHub} \\

    \textsc{Transfer} (\ref{S53})
    & \ding{70} & \ding{51} & ST & Java & \ding{55} & --
    & \href{https://github.com/soarsmu/transfer}{GitHub} \\

    \textsc{DirtyCred} (\ref{S54})
    & \ding{71} & \ding{55} & ST & C & \ding{51} & --
    & \href{https://github.com/Markakd/DirtyCred}{GitHub} \\

    \textsc{DualSC} (\ref{S55})
    & \ding{70} & \ding{55} & ST & Python & \ding{51} & --
    & \href{https://github.com/NTDXYG/DualSC}{GitHub} \\

    \textsc{TEFuzz} (\ref{S56})
    & \ding{70} & \ding{51} & ST & Python & \ding{51} & Docker
    & \href{https://github.com/seclab-fudan/TEFuzz}{GitHub} \\

    \textsc{ExGen} (\ref{S57})
    & \ding{70} & \ding{55} & RP & Python & \ding{51} & --
    & \href{https://tinyurl.com/mrzn5km3}{Link} \\

    \textsc{ExploitGen} (\ref{S58})
    & \ding{70} & \ding{55} & ST & Python & \ding{55} & --
    & \href{https://github.com/NTDXYG/ExploitGen}{GitHub} \\

    \textsc{RestTestGen} (\ref{S60})
    & \ding{70} & \ding{51} & ST & Java & \ding{51} & Docker
    & \href{https://github.com/SeUniVr/RestTestGen}{GitHub} \\

    \textsc{Rengar} (\ref{S61})
    & \ding{70} & \ding{51} & ST & Java & \ding{55} & --
    & \href{https://github.com/d1tto/Rengar}{GitHub} \\

    \textsc{APIDiff} (\ref{S62})
    & \ding{71} & \ding{55} & ST & NodeJS & \ding{51} & --
    & \href{https://github.com/OSUSecLab/APIDiff}{GitHub} \\

    \textsc{SMRL} (\ref{S63})
    & \ding{70} & \ding{51} & EF & Java & \ding{51} & --
    & \href{https://sntsvv.github.io/SMRL/}{Web} \\

    \textsc{Bagua} (\ref{S64})
    & \ding{71} & \ding{55} & RP & Python & \ding{51} & --
    & \href{https://github.com/Stab1el/BAGUA}{GitHub} \\

    \textsc{Vesta} (\ref{S67})
    & \ding{70} & \ding{51} & ST & Python & \ding{51} & Docker
    & \href{https://github.com/chen-zirui/TestMigration}{GitHub} \\

    \hline
    \multicolumn{8}{p{9.5cm}}{Docs $=$ Availability of documentation or README files. Deploy. $=$ Availability of deployment artifacts, such as Docker or virtual machine images.}
\end{tabular}
}
\end{table}

Of the \totalStudies studies, we found 31 (46\%) with a publicly accessible link (as of September 23, 2026) to the tool implementation.
Each tool has a given name, and most are hosted on \textsc{GitHub}, even though most did not receive regular maintenance after the first release.
The study of \textsc{Mista} (\ref{S10}) provides a non-working link, but we recovered the repository where the project was published through an online search.
The link provided by the study of \textsc{SSOScan} (\ref{S17}) instead redirects to an older snapshot of the same website hosted on \textsc{Internet Archive's Wayback Machine}.

In 6 studies, the paper provided a link; however, it no longer worked during our analysis, and we could not find a replacement online.
In 3 studies, the paper explains that the tool is available on request.
In the remaining 27 cases, the study did not mention the tool's availability.
We observed that not releasing the tool was more common in the early 2010s, but it still occurs in several cases after 2020.
We listed studies without an accessible tool in the paper's online appendix~\cite{replicationpackage}.

We classified the 31 publicly available tools as either \textit{fully automated} or \textit{semi-automated}, depending on whether some degree of user intervention is required during execution.
This classification excludes the manual actions required to prepare the input provided to the tool and to analyze its output.
We found a clear prevalence of fully automated solutions: 22 of 31 tools (71\%), while the remaining 9 (29\%) are semi-automated.
Thus, although most released tools can execute their main workflow automatically, not all of them can be used in fully automated assessment pipelines without developing additional harnesses or performing recurring manual activities.

Automation level and CI/CD suitability capture related but distinct properties.
The former describes whether the tool can execute its proposed workflow without user intervention once it has been launched.
The latter additionally considers whether this workflow can be integrated into a CI/CD pipeline with limited engineering effort.
Consequently, full automation is not sufficient for CI/CD suitability: a fully automated tool may still require extensive target-specific configuration, an interactive or fragile execution environment, expert interpretation of its output, or manual validation that is essential to determine whether the result is successful.

From a DevOps perspective, 14 tools (45\%) appear suitable for integration into CI/CD pipelines with limited additional engineering effort.
We considered a tool suitable when its main workflow could run unattended after any one-time target-specific configuration, accept programmatically supplied inputs, and automatically produce machine-consumable artifacts or results.
Tools requiring recurring manual intervention, expert analysis, interactive execution, or indispensable manual validation were considered not CI/CD-suitable.

For example, \textsc{TEFuzz} (\ref{S56}) automatically executes generated test cases, analyzes the resulting reports, and synthesizes exploit payloads.
Although it requires a target-specific driver and adaptation rules during the initial setup, its main workflow can subsequently run without human intervention and can therefore be integrated into a CI/CD pipeline with limited additional engineering effort.
In contrast, \textsc{Centaur} (\ref{S41}) requires specialist knowledge and target-specific Android symbolic-execution configuration, while its outputs are primarily intended for expert interpretation.
\textsc{Fugio} (\ref{S52}) lies between these cases: it automatically generates exploit payloads, but depends on an external runtime environment and manual validation.
These examples illustrate why automation level and CI/CD suitability should not be treated as equivalent: the former concerns the execution of the tool's workflow, whereas the latter also concerns reproducibility, integration effort, and the degree to which the results can be consumed automatically.

We grouped the 31 publicly available tool releases into three mutually exclusive categories: \textit{Standalone Tool}, \textit{Research Prototype}, and \textit{Extensible Framework}.

\begin{itemize}
    \item \textit{Standalone Tool} (18 tools, 58\%). A released implementation
    that provides a direct workflow for its task. It accepts the required target or configuration inputs and produces the corresponding results without requiring users to extend the implementation. It may nevertheless depend on third-party software or a target-specific execution environment.

    \item \textit{Research Prototype} (8 tools, 26\%). A released implementation whose primary purpose is to demonstrate or evaluate the technique proposed in the paper. Its operation remains substantially tied to research-specific assumptions, experimental inputs, specialist infrastructure, or manual setup, and it provides no explicit mechanism for extending it beyond the presented use case.

    \item \textit{Extensible Framework} (5 tools, 16\%). A released platform that provides explicit mechanisms for adding or adapting functionality beyond the original evaluation. Such mechanisms may include a domain-specific language, a public API, plug-in components, pluggable generators or analyses, or documented integration points.
\end{itemize}

When categories could overlap, we gave priority to explicit extensibility.
Among the remaining tools, implementations primarily presented as research-specific proof-of-concept demonstrations were classified as \textit{Research Prototypes}, while the others were classified as \textit{Standalone Tools}.

Python (16 cases, 52\%) and Java (11 cases, 36\%) are the dominant runtime platforms.
Most tools can therefore run with a Python interpreter or a Java platform, provided that the appropriate version is used.
The Python-based tools are often distributed as a collection of \texttt{.py} script files, whereas Java-based tools are commonly packaged as JARs.
In most cases, users have to select the appropriate runtime and download the required dependencies themselves.

Documentation or README files are available for 26 of the 31 tools (84\%).
However, the level of detail and completeness varies considerably, which may hinder reuse and reproduction.

Some tools also provide deployment support to facilitate reproducibility and practical adoption.
In particular, \textsc{Vesta} (\ref{S67}), \textsc{RestTestGen} (\ref{S60}), \textsc{TEFuzz} (\ref{S56}), \textsc{Fugio} (\ref{S52}), and \textsc{Savior} (\ref{S44}) provide Docker images, whereas \textsc{Rivulet} (\ref{S42}) distributes a pre-configured virtual machine image.
Overall, only 6 tools (19\%) are distributed with deployment artifacts.

\rqanswer{6}{Less than half of the reviewed studies (46\%) provide a publicly accessible tool implementing an APG approach, and almost one out of three of these tools (29\%) is not fully automated.
Fourteen tools (45\%) appear suitable for integration into CI/CD pipelines with limited additional engineering effort, although automation level and CI/CD suitability capture different properties.
Python (52\%) and Java (36\%) are the primary runtime platforms.
Documentation or README files are available for 84\% of the publicly available tools, whereas only 6 tools (19\%) are distributed with deployment artifacts such as Docker or virtual machine images.
Over half of the publicly available artifacts (58\%) are released as standalone tools, while 26\% are research prototypes and 16\% are extensible frameworks.
Despite the availability of source code for these tools, reproducibility and practical integration remain challenging because of target-specific setup requirements, heterogeneous runtime environments, and the need for manual validation or expert interpretation in several cases.
}

\section{Summary and Discussion}

Table~\ref{tab:rq-summary} summarizes the findings and recommendations from each research question.
Next, we discuss the existing limitations and challenges affecting the field.
We outline recommended actions that interested research communities should pursue to expand the knowledge in the field and discuss the practical implications for software developers who wish to adopt such tools in their day-to-day work.

\begin{table}[t]
\centering
\caption{Summary of findings and recommendations from each research question.}
\label{tab:rq-summary}
\resizebox{\linewidth}{!}{%
\begin{tabular}{|l|p{6.5cm}|p{6.5cm}|}
\hline
\textbf{Research Question} & \textbf{Key Findings} & \textbf{Recommendations} \\
\hline
\multirow{2}{4cm}{RQ$_1$ -- \textcolor{black!75}{\textit{``\rqOne''}}} & \faSearchPlus{} Current APG research often targets \textit{web-facing software} and \textit{command-line utilities}. This may be due to a tendency to draw inspiration from existing APG approaches and the possibility to directly compare with them. & \faLightbulb{} More focus on other domains is needed, such as \textit{mobile platforms} and \textit{smart contracts}, as well as involve domains like machine learning libraries~\cite{harzevili2023characterizing}. \\
\cline{2-3}
& \faSearchPlus{} Current APG research often targets memory-related and injection weaknesses, which are linked to C/C++ and PHP software, respectively. This may be due to the large availability of known vulnerabilities of this kind and their attribution to localized bugs. & \faLightbulb{} High-diversity datasets of compiled applications with known vulnerabilities should be developed. They should also involve more complex vulnerabilities, such as design-level issues~\cite{cawe:icsaw2017:santos}. \\
\hline
\multirow{2}{4cm}{RQ$_2$ -- \textcolor{black!75}{\textit{``\rqTwo''}}} & \faSearchPlus{} The probes generated by current APG research are not always \textit{self-verifying}. This may be due to the challenges in defining the expected/wrong behavior. & \faLightbulb{} More effort is needed in generating solid oracles, so as to increase the probe usefulness and the likelihood of using the APG tools in real-world settings. \\
\hline
\multirow{2}{4cm}{RQ$_3$ -- \textcolor{black!75}{\textit{``\rqThree''}}}
& \faSearchPlus{} Current APG research relies on a restricted set of established analysis tools such as \textsc{Angr}, \textsc{Klee}, and \textsc{Z3}. APG research may be bounded by the capabilities of available analysis tools. & \faLightbulb{} New foundational analysis tools are needed, especially if they have broad programming language coverage. \\
\hline
\multirow{2}{4cm}{RQ$_4$ -- \textcolor{black!75}{\textit{``\rqFour''}}} & \faSearchPlus{} Current APG tools require executable versions of complete programs/applications, and often they ask for complex auxiliary inputs. These provide better guidance to the generation algorithm, but the user has to invest more effort in preparing the required input. & \faLightbulb{} APG tools should be designed to work on incomplete programs/applications, as a full executable may come at later development stages (i.e., enforce the \textit{shift left} principle).
APG tools should also ask the user for minimal data to relieve their burden, and produce what they need on their own. \\
\hline
\multirow{2}{4cm}{RQ$_5$ -- \textcolor{black!75}{\textit{``\rqFive''}}} & \faSearchPlus{} Current APG research evaluates the effectiveness of the approaches on small real-world datasets. This may be due to a lack of standardized benchmarks. & \faLightbulb{} More effort is needed to develop benchmark platforms that allow the evaluation of multiple APG approaches on the same target set and with a common metric suite. \\
\hline
\multirow{2}{4cm}{RQ$_6$ -- \textcolor{black!75}{\textit{``\rqSix''}}} & \faSearchPlus{} Only a relatively small share of current APG tools may be theoretically included in CI/CD pipelines, but this has never been demonstrated. Current studies aim to present a research prototype. & \faLightbulb{} More priority should be given to boosting tool installation, usability, and seamless integration with automated pipelines to foster industry adoption. \\ 
\hline
\end{tabular}
}
\end{table}

\subsection{Triangulating Underexplored Areas}

The three most popular programming languages that APG approaches support are C/C++ (19 studies out of \totalStudies), Java (14 studies, including approaches targeting bytecode), and PHP (13 studies).
If we include language-agnostic approaches, these numbers rise to 27, 22, and 21, respectively.
While such languages remain among the top programming languages with reported vulnerabilities~\cite{veracode}, others also exhibit significant vulnerability counts, such as JavaScript and Python.
Unfortunately, only a small fraction of studies (a total of 6 out of \totalStudies) have covered such languages specifically (rising to 16 if we include language-agnostic approaches).
The result suggests limited research interest in such languages, despite their frequent exposure to security vulnerabilities.
This calls on researchers to invest attention and effort in \textbf{generating exploits and security tests for underexplored languages}, primarily JavaScript and Python.
An initial step could be to \textit{port} existing automated solutions to support other languages.
In general, it may be worth not binding to a single language but adopting \textit{language-agnostic} designs when developing new APG approaches, whenever possible.

At the same time, this is not as straightforward as it may seem.
Researchers may choose to target a specific programming language \textbf{as a consequence of the vulnerabilities and software types they focus on}.
From our review, we observed that studies often describe (in the initial sections) the research problem by deciding upfront the vulnerability types to exploit/test and the software type, thereby fixing the \textit{scope}.
This choice is often motivated by the relevance of the vulnerabilities and software types, as well as by prior research on the topic.
By contrast, the target programming language is part of the APG approach's design, serving more as a \textit{consequence} of the defined scope.
In this regard, we often find the vulnerability and software types mentioned in paper titles much more often than the programming language: 28 studies mention the vulnerability type, 28 studies mention the software type, 46 studies mention at least one of these two aspects, and only 5 studies mention the programming language.

Thus, porting existing APG approaches to another programming language is not straightforward and depends on whether the targeted context remains valid, especially with respect to the supported vulnerability types.
For example, system software is inevitably tied to C/C++, leaving little room to maneuver; however, memory-related vulnerabilities occur not only in C/C++ software but also in Rust, which could be considered an option for porting.
Similarly, web-based vulnerabilities often occur in JavaScript, which is widely used today for server-side login in web applications.

Furthermore, we cannot rule out the possibility that language choice is influenced by the availability of \textit{analysis tools} that APG approaches require to complete their tasks successfully.
As RQ$_3$ showed, there is a plethora of reusable tools for C/C++ code, e.g., symbolic execution engines and white-box fuzzers, that might be room to maneuver or even missing for other languages.

Lastly, we observed that other underexplored vulnerabilities and software types warrant attention.
This includes weaknesses such as \textit{unsafe deserialization} (CWE-502) and \textit{missing authentication} (CWE-306), both of which are listed in the 2024 CWE Top 25 Most Dangerous Software Weaknesses.
Unfortunately, these weaknesses are challenging to exploit and test in practice for different reasons.
Unsafe deserialization requires an accurate modeling of the application's \textit{attack surface}, which often includes deep object structures and potential gadget chains.
On the other hand, missing authentication is also difficult to detect automatically, as tools struggle to infer the intended \textit{access control policies} for the protected resources.
These challenges are further exacerbated by the use of \textit{third-party libraries}, which are a major channel for importing vulnerabilities from external sources~\cite{veracode,Wang:icsme20:library:risks,Zhan:icse21:third:party:vuln}.
Their inclusion expands the codebase that must be analyzed and introduces additional behavior that automated probe generators must account for.

\discussion{Covering underexplored research areas.}{
Researchers should devote themselves to covering less-considered programming languages, vulnerabilities, and software types.
For instance, developing automated probe generators targeting popular programming languages such as JavaScript and Python, other critical vulnerability types such as unsafe deserialization and authentication bugs, and software components such as third-party libraries.
}

\subsection{Challenges in Comparing the Generation Techniques}
\label{subsec:comparison}

All reviewed studies evaluated their APG approach on a certain dataset of test subjects (applications, programs, functions, regular expressions, etc.) and vulnerabilities to assess their effectiveness.
However, they rarely used the same datasets: all studies prepared their own evaluation sets, as shown in Table~\ref{tbl:rq5-evaluation}.
Two notable exceptions are \ref{S13}, which used the same set of programs and vulnerabilities as in \ref{S06} and \ref{S09}, and \ref{S56}, which used a subset of applications used by \ref{S16}.
Nevertheless, across the custom evaluation datasets, we observed several frequently used subjects, such as \textsc{WordPress}, \textsc{Joomla}, and \textsc{HotCRP} among the PHP web applications; \textsc{FFmpeg}, \textsc{FileZilla Server}, and \textsc{Chrome} among the C/C++ programs; and \textsc{HttpClient} as a Java library.
This decision is primarily driven by the knowledge that certain applications were vulnerable at specific versions, which is essential to verify that the exploit/test generator has at least one vulnerability to trigger.
Unfortunately, these recurring applications only cover a restricted set of common vulnerability types.
This reduces the generalizability of the observed performance, making it difficult to fairly compare different approaches without drawing wrong conclusions.
Thus, we argue for creating a \textbf{large, shared vulnerability dataset that enables fairer comparisons and more accurate assessments of APG effectiveness}.
Such a reference benchmark would help avoid validating approaches on \textit{cherry-picked} datasets, which can distort their true performance and make them appear to outperform competing approaches.

At the same time, we recognize that curating a representative collection of test subjects and vulnerabilities requires substantial effort, often beyond what is feasible within a research study focused on introducing a novel APG approach.
The main challenge lies in preparing the \textit{dynamic test environment} for all test subjects so that the generated exploits and security tests can be run against them without errors.
A first step toward this could be to gather all test subjects used in the reviewed studies (see Table~\ref{tbl:rq5-evaluation}) and prepare stable execution environments (i.e., sandboxes).
Subjects such as applications, programs, and functions could be stored in pre-compiled, isolated containers with the required OS version and dependencies installed, whereas regular expressions would primarily require a mechanism to feed them into the underlying pattern matcher.
However, this entire setup is complicated by the fact that less than half of the reviewed papers (31 out of \totalPapers) have publicly released their datasets, making it difficult to recreate the exact test environment they used.
Even worse, only 23 datasets are freely accessible (as of September 23, 2026).
Nevertheless, these datasets might not include all the files needed to reproduce the test environments correctly.

Lastly, the reviewed studies exhibit another type of discrepancy in their evaluation methodology: they often do not adopt a common definition of \textit{``success''} for generated exploits and security tests (some studies do not even provide a clear definition at all).
For example, \ref{S36} and \ref{S38} both generated exploits targeting Linux kernel vulnerabilities, but \ref{S36} counted success as the number of times a shell was spawned, whereas \ref{S38} relied on manual confirmation of exploitability.
These different definitions of success affect the resulting performance score and prevent meaningful comparison, even when using the same set of test subjects.
Recomputing metrics from raw results is often impossible without rerunning the tools---which can be challenging, as discussed in Section~\ref{subsec:reuse}.
This highlights the compelling need for a \textbf{shared definition of success and a standardized set of performance metrics} for APG tools.

\discussion{Enforce a Fairer Comparison Protocol.}{
Researchers should invest dedicated effort in developing a common reference dataset for comparing APG approaches.
The test subjects (functions, programs, applications, regular expressions, etc.) should have pre-built execution environments to facilitate benchmarking.
Studies introducing novel APG approaches should then refrain from using custom datasets to avoid bias and unrealistic performance assessment.
A clear definition of success should be adopted, enabling the use of standard performance metrics to assess the generator's effectiveness.
}

\subsection{Difficult (to Impossible) Reuse of Tools}
\label{subsec:reuse}

Another major challenge is that not all the reviewed papers (31 out of \totalPapers) publicly released the tools they introduced.
Only 31 of these are freely accessible (as of September 23, 2026), as listed in Table~\ref{tbl:avail-tools}.
As a result, external researchers cannot independently reuse these tools or reproduce the corresponding studies.
This situation raises several serious concerns. 
First, the veracity of the results reported in the original papers cannot be independently verified.
Second, non-runnable tools cannot be included in comparative analyses, making it difficult to assess whether novel APG approaches truly outperform the state of the art and forcing researchers to rely solely on the results disclosed by the original authors.
Third, extending or enhancing inaccessible tools is practically infeasible, as it would require re-implementing them from scratch based solely on the descriptions provided in the papers, which may not always be complete.
This could also explain why only a few papers (29 out of \totalPapers) included a direct comparison with other APG approaches.

In this systematic review, we did not assess the reproducibility of the tools, as this would require preparing the appropriate configuration and environment to run each tool according to its documentation.
A systematic reproducibility study would constitute a separate and substantial effort and may be pursued as future work stemming from this survey.
Nonetheless, we argue that researchers should be encouraged to \textbf{publicly release their tools and provide clear setup and replication instructions}. For instance, tools might be distributed as \textit{virtual machines} or \textit{containers} that encapsulate all required dependencies and system configurations, thereby facilitating reproducibility. Additionally, the community would benefit from the definition of common reporting guidelines, much like those proposed in closely related domains \cite{baltes2026guidelinesempiricalstudiessoftware,ralph2020empirical}.
These practices would enhance the study's credibility and increase its scientific impact.
Unfortunately, virtualization and containerization have been implemented in just 6 of the 31 (less than 20\%) accessible tools (as seen in Section~\ref{subsec:rq6}, indicating a need for additional engineering effort.
Without these, the only options left are to faithfully follow the steps described in the documentation (often in README files) or to contact the authors.

Nevertheless, the main challenges arise during dependency installation and proper configuration of the test environment, rather than in issuing commands to launch the tools.
Indeed, harnessing the test environment is likely the most complex part, especially given that several APG approaches rely on assumptions that are difficult to replicate outside the original study.
Thus, the tools should also integrate as many steps as possible to \textbf{automatically create the test environment}, i.e., the sandbox that isolates the test subjects and provides all required configurations.
Unlike the discussion point elaborated in Section~\ref{subsec:comparison}, the effort required here is predominantly technical rather than scientific.

\discussion{Facilitate Tool Reuse.}{
Researchers should release the APG tools they develop to encourage direct comparison and incremental improvements.
Wrapping tools in virtual machines or container images makes them even easier to install and run.
Besides, the test environment and the test subjects used in the evaluation should be pre-configured and released in sandboxes.
}

\subsection{The Real Goal of Exploits and Security Tests}

In Section~\ref{sec:intro}, we provided a convenience definition of \textit{probe} to encompass both \textit{exploits} and \textit{security tests}.
In our analyses, we observed that papers did not always clearly describe the type of artifacts generated or their actual level of ``weaponization'', i.e., how much harm they cause.
For instance, \ref{S16} states that it generates exploits when, in reality, they can only cause limited harm (i.e., they are PoC).
While this claim does not contradict what is claimed (indeed, a PoC is a type of exploit), this may leave readers confused about the real capability of that specific APG approach.

We argue that research work in this field should take additional care in clearly explaining (1) \textbf{the study goal}, i.e., whether to discover unknown vulnerabilities or assess the exploitability of known ones, and (2) the \textbf{harmfulness of the actions} carried out by the generated probe against the test subject.
Papers would benefit from briefly \textbf{introducing the terminology adopted} so as to avoid assuming a universally accepted meaning for the concepts of exploit and security tests, as these are often interpreted differently depending on the reader's background.

\discussion{Describe Artifact Goals.}{
Researchers should clearly describe the intended behavior and goals of the probes (exploits or security tests) generated by the proposed tools, including the extent of harm they cause to the test subject.
}

\subsection{The Missing Link to Industry}

Most evaluations of the proposed APG approaches were conducted on real-world software datasets (see Table~\ref{tbl:rq5-evaluation}); however, they were performed in controlled test environments rather than in actual deployment settings.
In particular, we observed that none of the reviewed papers validated or deployed the proposed APG approach in an industrial context.
We also found no quantifiable evidence of practitioners' successful adoption of these tools.
As indicated by the \textsc{GitHub} repositories of several tools (Table~\ref{tbl:avail-tools}), only a few of them received more than 100 stars, with \textsc{DirtyCred} \ref{S54} and \textsc{Fuze} \ref{S32} standing out with 593 and 501, respectively (as of September 23, 2026).
Still, there are no clear signals of adoption outside these research communities.
Hence, research should start by \textbf{investigating developers' inclinations and needs} to adopt APG tools, i.e., whether they would adopt one for certain parts of their activities.

A key question is whether these artifacts can be leveraged to detect potential \textit{security regressions}, i.e., the reintroduction of a previously fixed vulnerability.
Unfortunately, the exploits and security tests produced by the existing tools appear to be designed for one-off usage.
Future work should therefore provide solutions to \textbf{include these artifacts in existing project test suites} and mechanisms to \textbf{maintain them properly over time}.
This would make the generation tools a more valuable support for effective security testing.

Another step could be to \textbf{integrate the generation tools directly into CI/CD workflows} to enable continuous, automated production of exploits and security tests with every code change.
This would ensure that vulnerabilities are detected as early as possible in the development lifecycle, thereby strengthening \textit{``shift-left''} security practices.
Additionally, it further reduces the risk of regressions by automatically rechecking previously addressed weaknesses.
Prioritizing this integration is more relevant for developers than focusing solely on boosting the performance metrics, as done in the current research.
Our analysis for RQ$_6$ (Section~\ref{subsec:rq6}) indicates that 14 tools could, in principle, be integrated into CI/CD pipelines without substantial modification, as they do not appear to require user interaction.
However, a dedicated, empirical assessment is needed to confirm this happens in practice.

\discussion{Bridge Research Prototypes to Practice.}{
Researchers should investigate developers' inclinations and needs for tools that generate exploits and security tests.
Beyond performance and effectiveness, researchers should also work on transferring the research prototypes to industry by enabling (i) integration of the generated artifacts (exploits and security tests) into existing project test suites and (ii) incorporation of the generators into CI/CD workflows.
}

\subsection{Potential Misuses by Malicious Actors}

The actual harmful nature of exploits and security tests generated by the existing tools remains poorly understood.
In some cases, misuse appears theoretically possible, yet only five out of \totalPapers papers explicitly mentioned potential misuse or other ethical concerns.
Some of these tools could, in principle, be used to carry out harmful exploits against real software systems if properly configured.
This raises an important research dilemma concerning \textit{dual-use technologies}.
Tools that can automatically generate full RCE exploits would be highly valuable because they could help reveal and assess critical vulnerabilities; however, their misuse could lead to severe consequences.
Conversely, tools that generate only reachability tests or toy proofs of concept entail lower risk but are less useful for assessing the severity and impact of vulnerabilities.
Balancing ethical considerations with scientific progress is therefore challenging.

A first reasonable step is for research papers to \textbf{explicitly disclose potential risks of misuse and describe the safeguards adopted to prevent or mitigate them}.
Possible actions include restricting tools to operate solely in controlled test environments (e.g., sandboxes) and disallowing their use on real, publicly accessible software.
For example, by ensuring that testing tools targeting web applications cannot operate on public URLs.
However, these protection mechanisms may still be circumvented if the tools' source code is publicly released (a common practice for scientific reproducibility), as it provides malicious actors with a basis to re-adapt the tools for harmful purposes, e.g., disabling the safeguards.
This raises another dilemma: whether to release the full source code or only part of it, creating a trade-off between safety and transparency.

Another practice that researchers could regularly adopt is a \textbf{coordinated vulnerability disclosure} model.
If a generation tool discovers a new vulnerability (a zero-day), the vendor should be notified before the research results are first disseminated, allowing sufficient time to develop and release a patch before the vulnerability is disclosed or the tool becomes public.
Only a few of the reviewed papers explicitly notified the relevant stakeholders of the discovered zero-day vulnerability (though additional studies may have notified stakeholders regardless, even if this was not documented in the paper).
This approach still does not prevent malicious actors from attempting to exploit unpatched systems in the wild.

\discussion{Misuse Mitigation and Coordinated Disclosure.}{
Researchers should adopt countermeasures to prevent or mitigate potential misuse of APG tools.
If these tools also identify zero-day vulnerabilities, the relevant stakeholders should be notified in accordance with a coordinated disclosure model.
}

\section{Threats to Validity}
\label{sec:ttv}
The goal of our study was to provide a comprehensive overview of the current body of knowledge on automated probe (exploit and security test) generation.
This section discusses potential limitations that may have affected the comprehensiveness of our analysis, along with the methodological measures taken to mitigate them.
We specifically identified three primary sources of threats to validity: those related to \textit{literature selection}, those concerning the \textit{analysis and synthesis of the selected studies}, and those regarding the \textit{constructs} underlying our classification.

\paragraph{Literature Selection.}
One potential threat to validity arises from the literature selection process, as there is always a risk of incomplete coverage.
Although we employed a structured search strategy across multiple major academic databases, including Scopus, ACM Digital Library, IEEE Xplore, Web of Science, and DBLP, certain relevant studies may have been inadvertently excluded.
This could be due to limitations in the search string, as different studies might employ alternative terminology or domain-specific phrasing that was not explicitly captured by our query.
To mitigate this risk, we employed a search query that only included \textit{essential} terms, i.e., terms that are expected to be present in any relevant article.
In addition, we reviewed papers published over an extensive timeframe, from 2005 to 2024, to increase the number of papers returned after the initial search.
As observed in Section~\ref{sec:paper-collection}, we found that no publication goes back to 2008, hence increasing our confidence in the comprehensiveness of the search.
Additionally, we supplemented the database search with a snowballing approach, manually tracing citations and references from selected papers to identify additional works that might not have been retrieved through keyword-based queries.
Although this process helped improve coverage, our snowballing was limited to studies on automated exploit and security test generation, meaning that potentially relevant studies from adjacent domains may not have been included.
However, we justified this focus by aligning it with the research scope and the core objectives of this review.

An improper definition of inclusion/exclusion criteria might affect the final set of papers approved for review. To mitigate this risk, we conducted a calibration phase with four inspectors using a sample of 100 studies to refine the selection criteria for APG papers.
During this process, all 100 papers were inspected by reading their titles, abstracts, and keywords.
Then, each inspector proposed their opinion on whether the paper fit the scope and the reason why. All the reasons were collected and discussed together until an agreement on a common set of inclusion/exclusion criteria was reached.
Afterward, the papers discarded among the 100 were rapidly skimmed again to reassign the right criterion that led to their removal.

As an additional consideration, the databases considered for the search index a substantial portion of peer-reviewed research but restricted the corpus to high-impact venues; therefore, they may not comprehensively cover all security-related venues, particularly those associated with industry-driven research or emerging methodologies published in specialized security workshops.
In this respect, it is worth reporting that the goal of our study was to survey mature scientific contributions in the field rather than emerging trends or niche approaches.
As such, this potential limitation does not impact the scope of our investigation.

\paragraph{Literature Analysis and Synthesis.}
After selecting the relevant studies, we applied multiple manual analysis steps to classify APG approaches, assess automation levels, and evaluate tool availability. Given the diversity of methodologies and experimental settings reported in the literature, there is an inherent risk that some findings may be interpreted differently depending on the criteria used for classification.
The heterogeneity in research designs, evaluation benchmarks, and dataset sizes further complicates direct comparisons between studies, potentially affecting the validity of our conclusions.
This risk is more pronounced for judgment-intensive dimensions such as artifact maturity, oracle automation level, and the notion of ``success''.
To address these concerns, we adopted a structured data-extraction process in which multiple reviewers independently assessed each study to ensure consistency in categorization. Any discrepancies in classification or interpretation were resolved through discussion and consensus, reducing the likelihood that individual bias would influence the results. Additionally, we maintained transparency by explicitly reporting cases where information had to be inferred due to missing details in the original studies, and we released the review form and raw data in our supplementary materials to enable independent scrutiny.

A reliability threat is that papers were split into batches, each assigned to a single reviewer, so most fine-grained extraction was ultimately performed by one reviewer per study.
We mitigated this through the joint calibration phase, collective discussion of borderline cases, and cross-checking against the shared form, but we did not compute a formal inter-rater agreement metric (e.g., Cohen's or Fleiss' $\kappa$), which we consider a limitation.

\paragraph{Construct Validity.}
A related threat concerns the constructs underlying our classification: the definitions of exploit and security test (i.e., automated probe) were deliberately imposed rather than inherited from the primary studies, so our mapping of an approach's output may not always align with the original authors' framing. We mitigated this by classifying studies by the essence of the generated artifacts rather than by the terminology used.

\section{Conclusion}
\label{sec:conclusions}
Automated Probe Generation (APG), which includes both Automated Exploit Generation (AEG) and Automated Security Test Generation (ASTG), has emerged as a promising research area for improving vulnerability assessment and secure software development. Despite growing scientific interest, the field has evolved across multiple research communities that often adopt different terminology, assumptions, and evaluation practices, making it difficult to obtain a coherent view of the state of the art.
This systematic literature review addressed this gap by analyzing \totalPapers papers, grouped in \totalStudies studies, and providing a unified characterization of APG with respect to targeted software and vulnerabilities, generated artifacts, underlying techniques, required inputs, evaluation strategies, and tool support.

Our findings show that existing solutions are still largely specialized, typically targeting specific software domains or vulnerability classes, relying on heterogeneous assumptions regarding the generated artifacts and their validation, and offering varying degrees of automation and practical tool support. At the same time, we observe a growing convergence toward generation pipelines that integrate static analysis, symbolic reasoning, dynamic execution, search-based techniques, and, more recently, learning-based methods.

Our work also identifies several directions for future research. First, our research calls for shared terminology, standardized benchmarks, and common evaluation protocols to improve comparability and reproducibility across studies. Second, it calls for greater emphasis on generating artifacts that are directly reusable in practice, supported by reliable automated oracles and mature tool implementations. Finally, it highlights the need to extend APG approaches to modern software ecosystems, emerging vulnerability classes, and AI-assisted software engineering workflows, as well as to further evaluate them in industrial settings to support their practical adoption.

\begin{acks}
This work was partially supported by EU-funded project Sec4AI4Sec (grant no. 101120393) and funded by the Deutsche Forschungsgemeinschaft (DFG, German Research Foundation) -- 583980027 (SToCC).
\end{acks}

\section*{Data Availability}
The supplementary material is available on \textsc{Google Drive} and includes spreadsheets showing the papers at each stage of the selection process~\cite{replicationpackage}.
Additionally, the list of primary studies included in this review is accessible through our companion website: \url{https://apg-slr.github.io}.
Upon acceptance, the material will be archived on Zenodo, and the link will be replaced with the permanent DOI.

\section*{CRediT author statement}

\textbf{Quang-Cuong Bui}: Conceptualization, Methodology, Investigation, Data Curation, Writing - Original Draft, Visualization; 
\textbf{Emanuele Iannone}: Conceptualization, Methodology, Investigation, Data Curation, Writing - Original Draft, Writing - Review \& Editing, Supervision, Project administration, Funding acquisition; 
\textbf{Maria Camporese}: Methodology, Investigation, Data Curation, Writing - Original Draft, Visualization; 
\textbf{Torge Hinrichs}: Investigation, Data Curation; 
\textbf{Catherine Tony}: Investigation, Data Curation; 
\textbf{László Tóth}: Investigation, Data Curation; 
\textbf{Fabio Palomba}: Conceptualization, Methodology, Writing - Review \& Editing; 
\textbf{Péter Hegedűs}: Methodology, Writing - Review \& Editing; 
\textbf{Fabio Massacci}: Methodology, Writing - Review \& Editing, Supervision, Funding acquisition; 
\textbf{Riccardo Scandariato}: Conceptualization, Methodology, Writing - Review \& Editing, Supervision, Project administration, Funding acquisition

\bibliographystyle{ACM-Reference-Format}
\bibliography{references,included_studies}

\newpage
\appendix
\section{List of Reviewed Studies}
\label{appx:papers}
\begin{table}[h]
\centering
\footnotesize
\caption{List of reviewed studies. The first two columns present the study IDs and associated publications (one study may span multiple publications). The last two columns show the year of publication and the venues.}
\label{tbl:included-study-list}
\begin{tabular}{llll||llll}
\textbf{Study} & \textbf{Ref.} & \textbf{Year(s)} & \textbf{Venue(s)} & \textbf{Study} & \textbf{Ref.} & \textbf{Year(s)} & \textbf{Venue(s)} \\
\midrule
S01\labelstudy{S01} & \cite{Kayacik2005225} & 2005 & ACSAC & S35\labelstudy{S35} & \cite{Pewny2019111} & 2019 & EuroS\&P \\
S02\labelstudy{S02} & \cite{Brumley2008143} & 2008 & S\&P & S36\labelstudy{S36} & \cite{Wu20191187} & 2019 & USENIX-SEC \\
S03\labelstudy{S03} & \cite{DelGrosso20083125} & 2008 & COR & S37\labelstudy{S37} & \cite{Wei2019120152} & 2019 & IEEE Access  \\
S04\labelstudy{S04} & \cite{Kiezun2009199} & 2009 & ICSE & S38\labelstudy{S38} & \cite{Wang2019} & 2019 &  Cybersecurity \\
S05\labelstudy{S05} & \cite{Cai200927} & 2009 & S\&P & S39\labelstudy{S39} & \cite{Yilmaz2020386} & 2020 & ACSAC \\
S06\labelstudy{S06} & \cite{Avgerinos2011} & 2011 & NDSS & S40\labelstudy{S40} & \cite{Chen20201093} & 2020 & USENIX-SEC \\
S07\labelstudy{S07} & \cite{Babic201112} & 2011 & ISSTA & S41\labelstudy{S41} & \cite{Luo20202946} & 2020 & TMC \\
S08\labelstudy{S08} & \cite{Bisht2011575} & 2011 & CCS & S42\labelstudy{S42} & \cite{Hough2020284} & 2020 & ICSE \\
S09\labelstudy{S09} & \cite{Cha2012380} & 2012 & S\&P & S43\labelstudy{S43} & \cite{Liu2020286} & 2020 & ISSTA \\
S10\labelstudy{S10} & \cite{Xu2012526, Xu2015247} & 2012, 2015 & TDSC, TR & S44\labelstudy{S44} & \cite{Chen20201580} & 2020 & S\&P \\
S11\labelstudy{S11} & \cite{Cotroneo2013125} & 2013 & ASE & S45\labelstudy{S45} & \cite{Atlidakis2020387} & 2020 & ICST \\
S12\labelstudy{S12} & \cite{Tripp2013347} & 2013 & ISSTA & S46\labelstudy{S46} & \cite{Lee2020} & 2020 & NDSS \\
S13\labelstudy{S13} & \cite{Huang2014270} & 2014 & TR & S47\labelstudy{S47} & \cite{Muntean2021162} & 2021 & RAID \\
S14\labelstudy{S14} & \cite{Aydin2014193} & 2014 & ICST & S48\labelstudy{S48} & \cite{KallingalJoshy2021540} & 2021 & ISSTA \\
S15\labelstudy{S15} & \cite{Appelt2014259} & 2014 & ISSTA & S49\labelstudy{S49} & \cite{Liguori2021321} & 2021 & ISSRE \\
S16\labelstudy{S16} & \cite{Dahse201442} & 2014 & CCS & S50\labelstudy{S50} & \cite{Liu20211468} & 2021 & S\&P \\
S17\labelstudy{S17} & \cite{Zhou2014495} & 2014 & USENIX-SEC & S51\labelstudy{S51} & \cite{Turonova20224165} & 2022 & USENIX-SEC \\
S18\labelstudy{S18} & \cite{Hu2015177} & 2015 & USENIX-SEC & S52\labelstudy{S52} & \cite{Park2022197} & 2022 & USENIX-SEC \\
S19\labelstudy{S19} & \cite{Do201540} & 2015 & IST & S53\labelstudy{S53} & \cite{Kang2022276} & 2022 & ISSTA \\
S20\labelstudy{S20} & \cite{Eshete2015} & 2015 & NDSS & S54\labelstudy{S54} & \cite{Lin20221963} & 2022 & CCS \\
S21\labelstudy{S21} & \cite{Olivo2015616} & 2015 & CCS & S55\labelstudy{S55} & \cite{Yang2022361} & 2022 & SANER \\
S22\labelstudy{S22} & \cite{Alhuzali2016641, Alhuzali2018377} & 2016, 2018 & CCS, USENIX-SEC & S56\labelstudy{S56} & \cite{Zhao20233691} & 2023 & USENIX-SEC \\
S23\labelstudy{S23} & \cite{Jan201612, Jan2019335} & 2016, 2019 & ISSTA, TSE & S57\labelstudy{S57} & \cite{Jin2023650} & 2023 & TDSC \\
S24\labelstudy{S24} & \cite{Sudhodanan2016} & 2016 & NDSS & S58\labelstudy{S58} & \cite{Yang2023} & 2023 & JSS \\
S25\labelstudy{S25} & \cite{Wu2017283} & 2017 & ICSE & S59\labelstudy{S59} & \cite{Wang2023} & 2023 & COSE  \\
S26\labelstudy{S26} & \cite{Garcia2017661} & 2017 & ESEC/FSE & S60\labelstudy{S60} & \cite{Corradini20232553} & 2023 & ICSE \\
S27\labelstudy{S27} & \cite{Tang2017492} & 2017 & ICST & S61\labelstudy{S61} & \cite{Wang20232427} & 2023 & S\&P \\
S28\labelstudy{S28} & \cite{Yang2017143} & 2017 & RAID & S62\labelstudy{S62} & \cite{Wang20236629} & 2023 & USENIX-SEC \\
S29\labelstudy{S29} & \cite{Garmany2018300} & 2018 & ACSAC & S63\labelstudy{S63} & \cite{Chaleshtari20233430} & 2023 & TSE \\
S30\labelstudy{S30} & \cite{Yang2018} & 2018 & NDSS & S64\labelstudy{S64} & \cite{Li2023} & 2023 & NDSS \\
S31\labelstudy{S31} & \cite{Heelan2018763, Heelan20191689} & 2018, 2019 & SEC, CCS & S65\labelstudy{S65} & \cite{Xu2024293} & 2024 & TIFS  \\
S32\labelstudy{S32} & \cite{Wu2018781} & 2018 & SEC & S66\labelstudy{S66} & \cite{Yao20242852} & 2024 & TDSC \\
S33\labelstudy{S33} & \cite{Krupp20181317} & 2018 & SEC & S67\labelstudy{S67} & \cite{Chen20242820} & 2024 & ICSE \\
S34\labelstudy{S34} & \cite{Shen2018225} & 2018 & ASE \\
\bottomrule
\end{tabular}
\end{table}

\end{document}